\documentclass[10pt,a4paper]{article}
\usepackage{amsfonts,amsmath,amssymb,cite,graphicx}
\allowdisplaybreaks
\newcommand{\Imag}{\mathop{\textrm{Im}}}
\newcommand{\sgn}{\mathop{\textrm{sgn}}}

\begin{document}
\title{Zero-range potentials for Dirac particles: \\
Bound-state problems}
\author{Rados{\l}aw Szmytkowski \\*[3ex]
Faculty of Applied Physics and Mathematics,
Gda{\'n}sk University of Technology, \\
ul.\ Gabriela Narutowicza 11/12, 80--233 Gda{\'n}sk, Poland \\
Email: radoslaw.szmytkowski@pg.edu.pl}
\date{}
\maketitle
\begin{abstract} 
A model in which a Dirac particle in $\mathbb{R}^{3}$ is bound by
$N\geqslant1$ spatially distributed zero-range potentials is
presented. Interactions between the particle and the potentials are
modeled by subjecting a particle's bispinor wave function to certain
limiting conditions at the potential centers. Each of these conditions
is parametrized by a $2\times2$ Hermitian matrix (or, equivalently, a
real scalar and a real vector) and mixes the upper and the lower
components of the wave function. The problem of determining particle's
bound-state eigenenergies is reduced to the problem of finding real
zeroes of a determinant of a certain $2N\times2N$ matrix. As the lower
component of the particle's wave function is inverse-square singular
at each of the potential centers, the wave function itself is not
square-integrable. Nevertheless, one can define a scalar
pseudo-product with the property that wave functions belonging to
different eigenenergies are orthogonal with respect to it. The wave
functions may then be normalized so that their self-pseudo-products
are plus one, minus one or zero. An auxiliary set of Sturmian
functions is constructed and used to derive an explicit representation
of particle's matrix Green's function. For illustration purposes, two
particular systems are studied in detail:\ 1) a particle bound in a
field of a single zero-range potential, 2) a particle bound in a field
of two identical zero-range potentials. \\*[3ex]
\textbf{Keywords:} Dirac equation; zero-range potentials; 
contact interactions; point interactions \\*[1ex]
\end{abstract}
\maketitle
%
%\newpage
%
\section{Introduction}
\label{I}
\setcounter{equation}{0}
Useful information about some quantum mechanical systems may be
obtained by analyzing models in which interactions between their
constituents are of contact nature, i.e., occur only when distances
between the constituents are zero. In the literature, such idealized
models are often referred to as \mbox{contact-,} \mbox{point-} or
delta-interaction approximations. Their variants differ among
themselves in how the point interactions are mathematically built into
them. One procedure, employed mainly in analysis of one-dimensional
systems, is to use the Dirac delta function (and, occasionally, its
first spatial derivative) in the potential energy term in the
Hamiltonian of the system. Another possibility, used predominantly in
studies concerning two- and three-dimensional systems, is to impose
certain conditions on a wave function at those points in the
configuration space that correspond to the situation when two or more
components of the system are at the same location in the physical
space. It is the latter approach that underlies the so-called
``zero-range potential method'' (or ``zero-range potential
approximation''), which has been popularized by Demkov and Ostrovskii
\cite{Demk75} and by Drukarev \cite{Druk78}, and which is sometimes
used in theoretical atomic, molecular, and solid state physics.

The research conducted so far on the construction and use of
point-interaction models has focused mainly on nonrelativistic
systems. Only a relatively small number of works have dealt with
corresponding models for particles described by relativistic wave
equations; representative of that group are the publications listed in
Refs.\ \cite{Pere77,Cout90,Alon00,Albe05,Szmy05,Cout06,Szmy06,Calc14,
Pank14,Guil19}. In particular, in Refs.\ \cite{Szmy05,Szmy06} the
present author outlined a formalism that generalizes the
nonrelativistic one from Refs.\ \cite{Demk75,Druk78} and enables one
to study stationary scattering of Dirac particles off an arbitrary
system of spatially distributed point obstacles. The purpose of the
present work is to complement Refs.\ \cite{Szmy05,Szmy06} by
developing a model that allows one to consider Dirac particles bound
by a system of zero-range potentials in $\mathbb{R}^{3}$.

The paper is structured as follows. In Sec.\ \ref{II}, we present
basic principles of our model. The problem of orthogonality and
normalization of particle's bound-state eigenfunctions is discussed in
Sec.\ \ref{III}. An auxiliary set of Sturmian functions is introduced
in Sec.\ \ref{IV}. These functions are then used in Sec.\ \ref{V} to
construct an explicit representation of the Dirac--Green's function
associated with the problem. Two illustrative examples --- a particle
interacting with a single zero-range potential center and a particle
in a field of two identical zero-range potentials --- are worked out
in Sec.\ \ref{VI}. A brief discussion of possible further developments
of the model is provided in Sec.\ \ref{VII}.
%
%\newpage
%
\section{The model}
\label{II}
\setcounter{equation}{0}
We consider a Dirac particle of rest mass $m$, bound by a system of
$N\geqslant1$ zero-range potentials, located at the points
$\boldsymbol{r}_{n}$, $n=1,\ldots,N$. Everywhere in $\mathbb{R}^{3}$,
except at the potential centers, the time-independent bispinor wave
function $\Psi_{a}(\boldsymbol{r})$ describing the particle obeys the
Dirac equation
\begin{equation}
[-\mathrm{i}c\hbar\boldsymbol{\alpha}\cdot\boldsymbol{\nabla}
+mc^{2}\beta-E_{a}\mathcal{I}]\Psi_{a}(\boldsymbol{r})=0
\qquad (\mbox{$\boldsymbol{r}\neq\boldsymbol{r}_{n}$; 
$n=1,\ldots,N$}),
\label{2.1}
\end{equation}
where $\mathcal{I}$ is the unit $4\times4$ matrix,
$\boldsymbol{\alpha}$ and $\beta$ are the standard $4\times4$ Dirac
matrices \cite{Schi68}, while $E_{a}$ (assumed to be real and such
that $-mc^{2}<E_{a}\leqslant mc^{2}$) is particle's eigenenergy which
is to be determined. In the model that we propose in this work, the
wave function $\Psi_{a}(\boldsymbol{r})$ is taken in the form
\begin{subequations}
\begin{equation}
\Psi_{a}(\boldsymbol{r})
=\sum_{n=1}^{N}
\left(
\begin{array}{c}
f(k_{a}|\boldsymbol{r}-\boldsymbol{r}_{n}|)\chi_{an} \\*[1ex]
-\mathrm{i}\varepsilon_{a}k_{a}^{-1}\boldsymbol{\sigma}
\cdot\boldsymbol{\nabla}f(k_{a}|\boldsymbol{r}-\boldsymbol{r}_{n}|)
\chi_{an}
\end{array}
\right)
\label{2.2a}
\end{equation}
or equivalently
\begin{equation}
\Psi_{a}(\boldsymbol{r})
=\sum_{n=1}^{N}
\left(
\begin{array}{c}
f(k_{a}|\boldsymbol{r}-\boldsymbol{r}_{n}|)\chi_{an} \\*[1ex]
\mathrm{i}\varepsilon_{a}g(k_{a}|\boldsymbol{r}-\boldsymbol{r}_{n}|)
\boldsymbol{\mu}_{n}(\boldsymbol{r})\cdot\boldsymbol{\sigma}
\chi_{an}
\end{array}
\right).
\label{2.2b}
\end{equation}
\label{2.2}%
\end{subequations}
In Eqs.\ (\ref{2.2a}) and (\ref{2.2b}), and hereafter, the functions
$f(z)$ and $g(z)$ are defined to be\footnote{~In a two-dimensional
model, in which both the potential centers and the Dirac particle
itself are confined to the plane, the analogues of the elementary
functions $f(z)$ and $g(z)$ of Eqs.\ (\ref{2.3a}) and (\ref{2.3b})
will be the cylindrical Macdonald functions $K_{0}(z)$ and $K_{1}(z)$,
respectively.}
\begin{subequations}
\begin{equation}
f(z)=\frac{\mathrm{e}^{-z}}{z}
\label{2.3a}
\end{equation}
and
\begin{equation}
g(z)=-\frac{\mathrm{d}f(z)}{\mathrm{d}z}
=\frac{\mathrm{e}^{-z}}{z}+\frac{e^{-z}}{z^{2}},
\label{2.3b}
\end{equation}
\label{2.3}%
\end{subequations}
respectively, $k_{a}$ and $\varepsilon_{a}$ are eigenenergy-dependent
parameters defined as
\begin{subequations}
\begin{equation}
k_{a}=\frac{\sqrt{(mc^{2})^{2}-E_{a}^{2}}}{c\hbar},
\label{2.4a}
\end{equation}
\begin{equation}
\varepsilon_{a}=\sqrt{\frac{mc^{2}-E_{a}}{mc^{2}+E_{a}}},
\label{2.4b}
\end{equation}
\label{2.4}%
\end{subequations}
respectively, $\boldsymbol{\sigma}$ is the vector composed of the
Pauli matrices and
\begin{equation}
\boldsymbol{\mu}_{n}(\boldsymbol{r})
=\frac{\boldsymbol{r}-\boldsymbol{r}_{n}}
{|\boldsymbol{r}-\boldsymbol{r}_{n}|}
\label{2.5}
\end{equation}
is the unit vector pointing from the center $\boldsymbol{r}_{n}$ to
the observation point $\boldsymbol{r}$. The parameters $k_{a}$ and
$\varepsilon_{a}$ defined in Eqs.\ (\ref{2.4}) are easily seen to be
related through
\begin{subequations}
\begin{equation}
k_{a}=\frac{2mc}{\hbar}\frac{\varepsilon_{a}}{1+\varepsilon_{a}^{2}},
\label{2.6a}
\end{equation}
or conversely
\begin{equation}
\varepsilon_{a}
=\frac{1\mp\sqrt{1-(\hbar k_{a}/mc)^{2}}}{\hbar k_{a}/mc}
=\frac{\hbar k_{a}/mc}{1\pm\sqrt{1-(\hbar k_{a}/mc)^{2}}},
\label{2.6b}
\end{equation}
\label{2.6}%
\end{subequations}
with the upper (respectively, lower) signs chosen for $0\leqslant
E_{a}\leqslant mc^{2}$ (respectively, $-mc^{2}<E_{a}\leqslant0$). The
two-component spinors $\chi_{an}$, which are also to be determined,
may be interpreted as generalized superposition coefficients in the
linear combination (\ref{2.2}).

The selection of $f(z)$ in the form (\ref{2.3a}) guarantees that for
$\boldsymbol{r}\neq\boldsymbol{r}_{n}$ the $n$th term in the sum
(\ref{2.2a}) does solve the Dirac equation (\ref{2.1}) separately,
regardless of the particular choice of the spinor $\chi_{an}$ it
involves. At $\boldsymbol{r}=\boldsymbol{r}_{n}$, the upper and lower
components of that term exhibit the first- and the second-order
singularities, respectively. The reader may wish to observe that the
form of the upper component of $\Psi_{a}(\boldsymbol{r})$ in either of
Eqs.\ (\ref{2.2}) mimics that of a wave function used in the
nonrelativistic variant of the method \cite{Demk75,Druk78}, except
that in the present case the superposition coefficients are the Pauli
spinors rather than complex numbers.

\begin{figure}[t!]
\begin{center}
%\hspace*{-1em}
\includegraphics[width=0.75\textwidth]{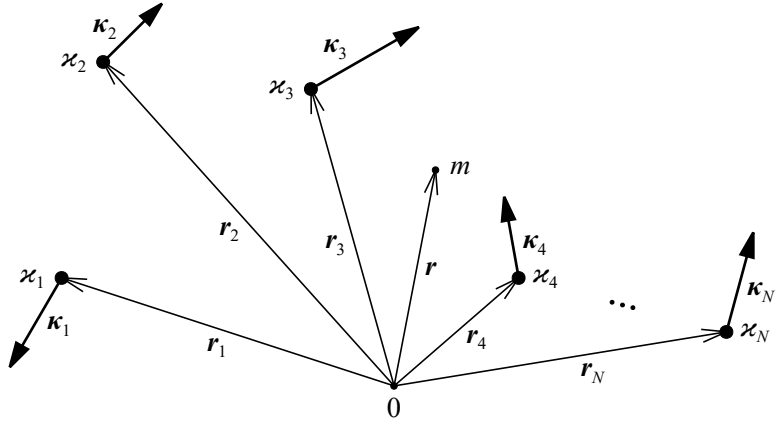}
\end{center}
\vspace*{-10ex}
\caption{The Dirac particle of mass $m$ and the position vector
$\boldsymbol{r}$ in the field of $N$ zero-range potentials. Each
potential center is characterized by its position vector
$\boldsymbol{r}_{n}$, the real scalar $\varkappa_{n}$ and the real
vector $\boldsymbol{\kappa}_{n}$. The relationship between the two
latter parameters and the $2\times2$ interaction matrix $K_{n}$
associated with the $n$th potential center is given in Eq.\
(\ref{2.10}).}
\label{FIG1}
\end{figure}

To complete our model, we represent the interaction between the Dirac
particle and the set of zero-range potentials by subjecting the
particle's wave function to the following limiting conditions:
\begin{equation}
\lim_{\boldsymbol{r}\to\boldsymbol{r}_{n}}
\left[\mathrm{i}(\boldsymbol{r}-\boldsymbol{r}_{n})
\cdot\boldsymbol{\alpha}^{(+)}
+\frac{\hbar}{2mc}|\boldsymbol{r}-\boldsymbol{r}_{n}|
\mathcal{K}_{n}^{(+)}
+\varepsilon_{a}k_{a}^{-1}\beta^{(+)}\right]\Psi_{a}(\boldsymbol{r})=0
\qquad (n=1,\ldots,N)
\label{2.7}
\end{equation}
at the points where the potentials are placed. Here, the $4\times4$
matrices $\boldsymbol{\alpha}^{(+)}$ and $\beta^{(+)}$, and their
counterparts $\boldsymbol{\alpha}^{(-)}$ and $\beta^{(-)}$ to be used
later, are defined to be
\begin{subequations}
\begin{equation}
\boldsymbol{\alpha}^{(\pm)}=\beta^{(\pm)}\boldsymbol{\alpha}
=\boldsymbol{\alpha}\beta^{(\mp)},
\label{2.8a}
\end{equation}
\begin{equation}
\beta^{(\pm)}=\frac{1}{2}(\mathcal{I}\pm\beta),
\label{2.8b}
\end{equation}
\label{2.8}%
\end{subequations}
respectively, while $\mathcal{K}_{n}^{(+)}$ are $4\times4$
energy-independent matrices that may be expressed in terms of the
$2\times2$ Hermitian matrices $K_{n}$ as
\begin{equation}
\mathcal{K}_{n}^{(+)}
=\left(
\begin{array}{cc}
K_{n} & 0 \\
0 & 0
\end{array}
\right),
\label{2.9}
\end{equation}
with zeroes denoting the $2\times2$ null matrices. In later
considerations we shall be exploiting the well-known fact that in the
Pauli basis, consisting of the unit $2\times2$ matrix $I$ and the
Pauli matrix vector $\boldsymbol{\sigma}$, the matrices $K_{n}$ have
the representations
\begin{equation}
K_{n}=\varkappa_{n}I+\boldsymbol{\kappa}_{n}\cdot\boldsymbol{\sigma},
\label{2.10}
\end{equation}
where
\begin{subequations}
\begin{equation}
\varkappa_{n}=\frac{1}{2}\textrm{Tr}\,K_{n},
\label{2.11a}
\end{equation}
and
\begin{equation}
\boldsymbol{\kappa}_{n}
=\frac{1}{2}\textrm{Tr}\,(\boldsymbol{\sigma}K_{n}).
\label{2.11b}
\end{equation}
\label{2.11}%
\end{subequations}
Since the matrices $K_{n}$ are presupposed to be Hermitian, the
scalars $\varkappa_{n}$ and the vectors $\boldsymbol{\kappa}_{n}$ are
real.

The limiting conditions (\ref{2.7}) guarantee (cf.\ Appendix) that
none of the points $\boldsymbol{r}_{n}$ is a source or a sink, i.e.,
that the flux across a spherical surface $\mathcal{S}_{n}$, centered
at the point $\boldsymbol{r}_{n}$ and of radius $\rho\to0$, does
vanish:
\begin{equation}
\lim_{\rho\to0}
\oint_{\mathcal{S}_{n}}\mathrm{d}^{2}\boldsymbol{\rho}_{n}\:
\boldsymbol{\mu}_{n}(\boldsymbol{r}_{n}+\boldsymbol{\rho}_{n})
\cdot\boldsymbol{j}_{a}(\boldsymbol{r}_{n}+\boldsymbol{\rho}_{n})=0
\qquad (n=1,\ldots,N),
\label{2.12}
\end{equation}
where $\boldsymbol{\rho}_{n}$ (with $|\boldsymbol{\rho}_{n}|=\rho$) is
the radius vector (respective to the center $\boldsymbol{r}_{n}$) for
a point on $\mathcal{S}_{n}$,
$\boldsymbol{\mu}_{n}(\boldsymbol{r}_{n}+\boldsymbol{\rho}_{n})
=\boldsymbol{\rho}_{n}/\rho$ [cf.\ Eq.\ (\ref{2.5})] is the unit
outward vector normal to $\mathcal{S}_{n}$ and
\begin{equation}
\boldsymbol{j}_{a}(\boldsymbol{r})=c\Psi_{a}^{\dag}(\boldsymbol{r})
\boldsymbol{\alpha}\Psi_{a}(\boldsymbol{r})
\label{2.13}
\end{equation}
(here and hereafter, the dagger denotes the Hermitian adjoint matrix)
is the Dirac current density vector.

Substitution of the wave function $\Psi_{a}(\boldsymbol{r})$ in the
form (\ref{2.2b}) into the limiting conditions (\ref{2.7}) yields the
following homogeneous algebraic system for the spinor superposition
coefficients $\chi_{an}$:
\begin{equation}
\left(\frac{\hbar}{2mc\varepsilon_{a}}K_{n}-I\right)\chi_{an}
+\sum_{\substack{n'=1 \\ (n'\neq n)}}^{N}
f(k_{a}|\boldsymbol{r}_{n}-\boldsymbol{r}_{n'}|)\chi_{an'}=0
\qquad (n=1,\ldots,N).
\label{2.14}
\end{equation}
With both current and future applications in mind, we introduce a
$2N\times2N$ matrix $\mathsf{L}(E)$ built of $2\times2$ blocks
\begin{equation}
L_{nn'}(E)
=\delta_{nn'}\left(\frac{\hbar}{2mc\varepsilon}K_{n}-I\right)
+(1-\delta_{nn'})f(k|\boldsymbol{r}_{n}-\boldsymbol{r}_{n'}|)I
\qquad (n,n'=1,\ldots,N),
\label{2.15}
\end{equation}
where $k$ and $\varepsilon$ are defined in terms of an energy
parameter $-mc^{2}<E\leqslant mc^{2}$ (which may or may not be equal
to one of particle's eigenenergies $E_{a}$) as [cf.\ Eqs.\
(\ref{2.4})]
\begin{subequations}
\begin{equation}
k=\frac{\sqrt{(mc^{2})^{2}-E^{2}}}{c\hbar}
\label{2.16a}
\end{equation}
and
\begin{equation}
\varepsilon=\sqrt{\frac{mc^{2}-E}{mc^{2}+E}},
\label{2.16b}
\end{equation}
\label{2.16}%
\end{subequations}
respectively. Defining also a $2N$-element column vector
\begin{equation}
\mathsf{x}_{a}
=\left(
\begin{array}{ccc}
\chi_{a1}^\mathrm{T} & \cdots & \chi_{aN}^\mathrm{T}
\end{array}
\right)^{\mathrm{T}}
\label{2.17} 
\end{equation}
(here and hereafter, T denotes the transpose matrix), we may rewrite
the system (\ref{2.14}) in the compact form
\begin{equation}
\mathsf{L}(E_{a})\mathsf{x}_{a}=\mathsf{0}.
\label{2.18}
\end{equation}
The system (\ref{2.18}) has nontrivial solutions $\mathsf{x}_{a}$ only
if the determinant of its matrix $\mathsf{L}(E_{a})$ vanishes:
\begin{equation}
\det\mathsf{L}(E_{a})=0.
\label{2.19}
\end{equation}
This is an algebraic equation for $E_{a}$ and its roots obeying the
constraint $-mc^{2}<E_{a}\leqslant mc^{2}$ play the role of particle's
bound-state eigenenergies in our model.

From Eqs.\ (\ref{2.14}) and (\ref{2.15}), one may deduce several
useful sesquilinear identities involving the spinor coefficients
$\chi_{an}$. We list four of them here. The first one is
\begin{equation}
\frac{\hbar^{2}}{2m}\sum_{n=1}^{N}\chi_{an}^{\dag}K_{n}\chi_{an}
+E_{a}k_{a}^{-1}\sum_{n,n'=1}^{N}
\mathrm{e}^{-k_{a}|\boldsymbol{r}_{n}-\boldsymbol{r}_{n'}|}
\chi_{an}^{\dag}\chi_{an'}
=c^{2}\hbar^{2}k_{a}\sum_{n,n'=1}^{N}\chi_{an}^{\dag}
\left[\frac{\partial L_{nn'}(E)}{\partial E}\right]_{E=E_{a}}
\chi_{an'}.
\label{2.20}
\end{equation}
It will find an application in Sec.\ \ref{III}, in the context of
normalization of the eigenfunctions (\ref{2.2}). The second one,
\begin{align}
& \hspace*{-2em}
\frac{E_{b}-E_{a}}{2mc^{2}}
\sum_{n=1}^{N}\chi_{bn}^{\dag}K_{n}\chi_{an}
-(k_{b}-k_{a})\sum_{n=1}^{N}\chi_{bn}^{\dag}\chi_{an}
\nonumber \\*
& \hspace*{3em}
+\sum_{\substack{n,n'=1 \\ (n\neq n')}}^{N}
\chi_{bn}^{\dag}\chi_{an'}
\big[k_{b}f(k_{b}|\boldsymbol{r}_{n}-\boldsymbol{r}_{n'}|)
-k_{a}f(k_{a}|\boldsymbol{r}_{n}-\boldsymbol{r}_{n'}|)\big]=0
\qquad (E_{b}\neq E_{a}),
\label{2.21}
\end{align}
may be shown to be closely linked to the orthogonality relation
(\ref{3.14}). The last two identities,
\begin{subequations}
\begin{align}
& (\varepsilon_{b}-\varepsilon_{a})
\sum_{n=1}^{N}\chi_{bn}^{\dag}\chi_{an}
-\sum_{\substack{n,n'=1 \\ (n\neq n')}}^{N}\chi_{bn}^{\dag}\chi_{an'}
\big[\varepsilon_{b}f(k_{b}|\boldsymbol{r}_{n}-\boldsymbol{r}_{n'}|)
-\varepsilon_{a}f(k_{a}|\boldsymbol{r}_{n}-\boldsymbol{r}_{n'}|)\big]
=0
\qquad (E_{b}\neq E_{a})
\nonumber \\*
& \hspace*{25em}
\label{2.22a}
\end{align}
and
\begin{align}
\frac{\hbar}{2mc}
\left(\varepsilon_{b}^{-1}-\varepsilon_{a}^{-1}\right)
\sum_{n=1}^{N}\chi_{bn}^{\dag}K_{n}\chi_{an}
+\sum_{\substack{n,n'=1 \\ (n\neq n')}}^{N}\chi_{bn}^{\dag}\chi_{an'}
\big[f(k_{b}|\boldsymbol{r}_{n}-\boldsymbol{r}_{n'}|)
-f(k_{a}|\boldsymbol{r}_{n}-\boldsymbol{r}_{n'}|)\big]=0
\qquad(E_{b}\neq E_{a}),
\label{2.22b}
\end{align}
\label{2.22}%
\end{subequations}
are presented here because of their structural simplicity.
%
%\newpage
%
\section{Orthogonality and normalization of eigenfunctions}
\label{III}
\setcounter{equation}{0}
Consider two bound-state eigenfunctions $\Psi_{a}(\boldsymbol{r})$ and
$\Psi_{b}(\boldsymbol{r})$, belonging to the energy eigenvalues
$E_{a}$ and $E_{b}$, respectively. If we premultiply the Dirac
equation obeyed by $\Psi_{a}(\boldsymbol{r})$ with
$\Psi_{b}^{\dag}(\boldsymbol{r})$ and integrate the result with
respect to $\boldsymbol{r}$ over the domain
\begin{equation}
\mathbb{R}_{\rho}^{3}
=\mathbb{R}^{3}\setminus\bigcup\limits_{n=1}^{N}\mathcal{V}_{n},
\label{3.1}
\end{equation}
where $\mathcal{V}_{n}$, $n=1,\ldots,N$, is a sphere of radius
\begin{equation}
0<\rho<\min_{1\leqslant n'\neq n''\leqslant N}
|\boldsymbol{r}_{n'}-\boldsymbol{r}_{n''}|
\label{3.2}
\end{equation}
centered at $\boldsymbol{r}_{n}$ (for simplicity, we choose radii of
all spheres $\mathcal{V}_{n}$ to be identical), this gives
\begin{equation}
\int_{\mathbb{R}_{\rho}^{3}}\mathrm{d}^{3}\boldsymbol{r}\:
\Psi_{b}^{\dag}(\boldsymbol{r})[\mathcal{H}-E_{a}\mathcal{I}]
\Psi_{a}(\boldsymbol{r})=0,
\label{3.3}
\end{equation}
where we have denoted
\begin{equation}
\mathcal{H}
=-\mathrm{i}c\hbar\boldsymbol{\alpha}\cdot\boldsymbol{\nabla}
+mc^{2}\beta.
\label{3.4}
\end{equation}
Integrating in Eq.\ (\ref{3.3}) by parts and exploiting the Gauss
divergence theorem yields
\begin{align}
& \hspace*{-5em}
\int_{\mathbb{R}_{\rho}^{3}}\mathrm{d}^{3}\boldsymbol{r}\:
\left\{[\mathcal{H}-E_{a}\mathcal{I}]
\Psi_{b}(\boldsymbol{r})\right\}^{\dag}\Psi_{a}(\boldsymbol{r})
-c\hbar\oint_{\mathcal{S}_{\infty}}
\mathrm{d}^{2}\boldsymbol{\rho}_{\infty}\:
\Psi_{b}^{\dag}(\boldsymbol{\rho}_{\infty})
\mathrm{i}\boldsymbol{n}_{\infty}\cdot\boldsymbol{\alpha}
\Psi_{a}(\boldsymbol{\rho}_{\infty})
\nonumber \\
& +\frac{c\hbar}{\rho}\sum_{n=1}^{N}
\oint_{\mathcal{S}_{n}}\mathrm{d}^{2}\boldsymbol{\rho}_{n}\:
\Psi_{b}^{\dag}(\boldsymbol{r}_{n}+\boldsymbol{\rho}_{n})
\mathrm{i}\boldsymbol{\rho}_{n}\cdot\boldsymbol{\alpha}
\Psi_{a}(\boldsymbol{r}_{n}+\boldsymbol{\rho}_{n})=0,
\label{3.5}
\end{align}
where $\boldsymbol{n}_{\infty}$ is the outward unit vector on the
spherical surface at infinity ($\mathcal{S}_{\infty}$) at the point
characterized by the (infinite) radius vector
$\boldsymbol{\rho}_{\infty}$, while $\boldsymbol{\rho}_{n}$ (with
$|\boldsymbol{\rho}_{n}|=\rho$) is the radius vector (respective to
the center $\boldsymbol{r}_{n}$) for a point on the spherical surface
$\mathcal{S}_{n}$ bounding $\mathcal{V}_{n}$ [cf.\ the comments
following Eq.\ (\ref{2.12})]. We observe that because both
eigenfunctions $\Psi_{a}(\boldsymbol{r})$ and
$\Psi_{b}(\boldsymbol{r})$ decay exponentially for $r\to\infty$, the
surface integral over $\mathcal{S}_{\infty}$ vanishes. As regards the
surface integrals over $\mathcal{S}_{n}$, with the use of the identity
\begin{equation}
\boldsymbol{\alpha}=\boldsymbol{\alpha}^{(+)}
+\boldsymbol{\alpha}^{(-)}
\label{3.6}
\end{equation}
we split each of them into two integrals and then modify the one
containing the matrix $\boldsymbol{\alpha}^{(-)}$ using
\begin{equation}
\boldsymbol{\alpha}^{(-)}=\boldsymbol{\alpha}^{(+)\dag}.
\label{3.7}
\end{equation}
This converts Eq.\ (\ref{3.5}) into
\begin{align}
& \hspace*{-5em}
(E_{b}-E_{a})\int_{\mathbb{R}_{\rho}^{3}}
\mathrm{d}^{3}\boldsymbol{r}\:
\Psi_{b}^{\dag}(\boldsymbol{r})\Psi_{a}(\boldsymbol{r})
+\frac{c\hbar}{\rho}\sum_{n=1}^{N}
\oint_{\mathcal{S}_{n}}\mathrm{d}^{2}\boldsymbol{\rho}_{n}\:
\Psi_{b}^{\dag}(\boldsymbol{r}_{n}+\boldsymbol{\rho}_{n})
\mathrm{i}\boldsymbol{\rho}_{n}\cdot\boldsymbol{\alpha}^{(+)}
\Psi_{a}(\boldsymbol{r}_{n}+\boldsymbol{\rho}_{n})
\nonumber \\*
& -\frac{c\hbar}{\rho}\sum_{n=1}^{N}
\oint_{\mathcal{S}_{n}}\mathrm{d}^{2}\boldsymbol{\rho}_{n}\:
[\mathrm{i}\boldsymbol{\rho}_{n}\cdot\boldsymbol{\alpha}^{(+)}
\Psi_{b}(\boldsymbol{r}_{n}+\boldsymbol{\rho}_{n})]^{\dag}
\Psi_{a}(\boldsymbol{r}_{n}+\boldsymbol{\rho}_{n})=0.
\label{3.8}
\end{align}
So far, the radius $\rho$ has been arbitrary except for being
subjected to the constraint (\ref{3.2}). At this stage, we let it tend
to zero. Transforming the surface integrals over $\mathcal{S}_{n}$
with the aid of the limiting conditions (\ref{2.7}) and making use of
the fact that
\begin{equation}
\varepsilon_{b}k_{b}^{-1}-\varepsilon_{a}k_{a}^{-1}
=-\frac{E_{b}-E_{a}}{c\hbar}
(\varepsilon_{b}k_{b}^{-1})(\varepsilon_{a}k_{a}^{-1})
\label{3.9}
\end{equation}
casts Eq.\ (\ref{3.8}) into the form
\begin{align}
& \hspace*{-5em}
(E_{b}-E_{a})\lim_{\rho\to0}
\Bigg\{\int_{\mathbb{R}_{\rho}^{3}}\mathrm{d}^{3}\boldsymbol{r}\:
\Psi_{b}^{\dag}(\boldsymbol{r})\Psi_{a}(\boldsymbol{r})
\nonumber\\
& -\frac{(\varepsilon_{b}k_{b}^{-1})
(\varepsilon_{a}k_{a}^{-1})}{\rho}\sum_{n=1}^{N}
\oint_{\mathcal{S}_{n}}\mathrm{d}^{2}\boldsymbol{\rho}_{n}\:
\Psi_{b}^{\dag}(\boldsymbol{r}_{n}+\boldsymbol{\rho}_{n})\beta^{(+)}
\Psi_{a}(\boldsymbol{r}_{n}+\boldsymbol{\rho}_{n})\Bigg\}=0.
\label{3.10}
\end{align}
From Eq.\ (\ref{3.10}), we see that if $E_{a}$ and $E_{b}$ are
distinct, then the eigenfunctions $\Psi_{a}(\boldsymbol{r})$ and
$\Psi_{b}(\boldsymbol{r})$ are orthogonal in the sense of
\begin{align}
& \lim_{\rho\to0}\left\{\int_{\mathbb{R}_{\rho}^{3}}
\mathrm{d}^{3}\boldsymbol{r}\:
\Psi_{b}^{\dag}(\boldsymbol{r})\Psi_{a}(\boldsymbol{r})
-\frac{(\varepsilon_{b}k_{b}^{-1})
(\varepsilon_{a}k_{a}^{-1})}{\rho}\sum_{n=1}^{N}
\oint_{\mathcal{S}_{n}}\mathrm{d}^{2}\boldsymbol{\rho}_{n}\:
\Psi_{b}^{\dag}(\boldsymbol{r}_{n}+\boldsymbol{\rho}_{n})\beta^{(+)}
\Psi_{a}(\boldsymbol{r}_{n}+\boldsymbol{\rho}_{n})\right\}=0
\nonumber \\*
& \hspace*{30em} (E_{a}\neq E_{b}).
\label{3.11}
\end{align}

If for any two sufficiently regular four-component functions
$\Phi(\boldsymbol{r})$ and $\Phi^{\prime}(\boldsymbol{r})$ we define
their volume
\begin{equation}
\big<\Phi\big|\Phi'\big>_{\mathbb{R}_{\rho}^{3}}
\equiv\int_{\mathbb{R}_{\rho}^{3}}\mathrm{d}^{3}\boldsymbol{r}\:
\Phi^{\dag}(\boldsymbol{r})\Phi^{\prime}(\boldsymbol{r})
\label{3.12}
\end{equation}
and surface
\begin{equation}
\big(\Phi\big|\Phi'\big)_{\mathcal{S}_{n}}
=\oint_{\mathcal{S}_{n}}\mathrm{d}^{2}\boldsymbol{\rho}_{n}\:
\Phi^{\dag}(\boldsymbol{r}_{n}+\boldsymbol{\rho}_{n})
\Phi^{\prime}(\boldsymbol{r}_{n}+\boldsymbol{\rho}_{n})
\label{3.13}
\end{equation}
scalar products, then the orthogonality relation (\ref{3.11}) may be
compactly rewritten as
\begin{equation}
\lim_{\rho\to0}
\left\{\big<\Psi_{b}\big|\Psi_{a}\big>_{\mathbb{R}_{\rho}^{3}}
-\frac{(\varepsilon_{b}k_{b}^{-1})
(\varepsilon_{a}k_{a}^{-1})}{\rho}
\sum_{n=1}^{N}\big(\Psi_{b}\big|
\beta^{(+)}\Psi_{a}\big)_{\mathcal{S}_{n}}\right\}=0
\qquad (E_{a}\neq E_{b}).
\label{3.14}
\end{equation}
In what follows, we shall be assuming that eigenfunctions belonging to
degenerate energy eigenvalues (if there are any) have been linearly
transformed among themselves so that the orthogonality relation
\begin{equation}
\lim_{\rho\to0}
\left\{\big<\Psi_{b}\big|\Psi_{a}\big>_{\mathbb{R}_{\rho}^{3}}
-\frac{(\varepsilon_{b}k_{b}^{-1})
(\varepsilon_{a}k_{a}^{-1})}{\rho}
\sum_{n=1}^{N}\big(\Psi_{b}\big|
\beta^{(+)}\Psi_{a}\big)_{\mathcal{S}_{n}}\right\}=0
\qquad (a\neq b)
\label{3.15}
\end{equation}
holds as long as $\Psi_{a}(\boldsymbol{r})$ and
$\Psi_{b}(\boldsymbol{r})$ are linearly independent, even if
$E_{a}=E_{b}$ (i.e., if $\varepsilon_{a}=\varepsilon_{b}$ and
$k_{a}=k_{b}$).

One may look on the sesquilinear form
\begin{equation}
\langle\!\langle\Psi_{b}\big|\Psi_{a}\rangle\!\rangle
\stackrel{\mathrm{def}}{=}\lim_{\rho\to0}
\left\{\big<\Psi_{b}\big|\Psi_{a}\big>_{\mathbb{R}_{\rho}^{3}}
-\frac{(\varepsilon_{b}k_{b}^{-1})
(\varepsilon_{a}k_{a}^{-1})}{\rho}
\sum_{n=1}^{N}\big(\Psi_{b}\big|
\beta^{(+)}\Psi_{a}\big)_{\mathcal{S}_{n}}\right\}
\label{3.16}
\end{equation}
as a scalar pseudo-product of two eigenfunctions
$\Psi_{b}(\boldsymbol{r})$ and $\Psi_{a}(\boldsymbol{r})$. It is not
sign-definite since in the product of $\Psi_{a}(\boldsymbol{r})$ with
itself, i.e., in the form
\begin{equation}
\langle\!\langle\Psi_{a}\big|\Psi_{a}\rangle\!\rangle
=\lim_{\rho\to0}
\left\{\big<\Psi_{a}\big|\Psi_{a}\big>_{\mathbb{R}_{\rho}^{3}}
-\frac{\varepsilon_{a}^{2}k_{a}^{-2}}{\rho}
\sum_{n=1}^{N}\big(\Psi_{a}\big|
\beta^{(+)}\Psi_{a}\big)_{\mathcal{S}_{n}}\right\},
\label{3.17}
\end{equation} 
the expression between the curly brackets is a difference of two
nonnegative terms and nothing can be said a priori about the sign of
its limit. We define the pseudo-norm $||\Psi_{a}||$ of the
eigenfunction $\Psi_{a}(\boldsymbol{r})$ as
\begin{equation}
||\Psi_{a}||
=\sqrt{|\langle\!\langle\Psi_{a}\big|\Psi_{a}\rangle\!\rangle|}
\geqslant0.
\label{3.18}
\end{equation}
The eigenfunctions with vanishing pseudo-norm, i.e., those for which
it holds that
\begin{equation}
\langle\!\langle\Psi_{a}\big|\Psi_{a}\rangle\!\rangle=0,
\label{3.19}
\end{equation}
will be called null-eigenfunctions. In addition, we define the
signature $\Delta_{a}\in\{0,\pm1\}$ of the eigenfunction
$\Psi_{a}(\boldsymbol{r})$ as
\begin{equation}
\Delta_{a}
=\sgn\langle\!\langle\Psi_{a}\big|\Psi_{a}\rangle\!\rangle
\label{3.20}
\end{equation}
(please observe that null-eigenfunctions have signature zero).

In standard quantum mechanics, it is frequently convenient to work
with bound-state eigenfunctions normalized to the unity with respect
to the natural norm induced by a scalar product under which
eigenfunctions belonging to different eigenvalues are orthogonal.
Provided that $\Delta_{a}=\pm1$, in our case the analogous normalizing
role is played by the constraint
\begin{equation}
\langle\!\langle\Psi_{a}\big|\Psi_{a}\rangle\!\rangle=\Delta_{a},
\label{3.21}
\end{equation}
which, at least formally, determines $\Psi_{a}(\boldsymbol{r})$ up to
a multiplicative phase factor. On combining Eq.\ (\ref{3.21}) with the
orthogonality constraint (\ref{3.15}), we then obtain the generalized
orthonormality relation
\begin{equation}
\langle\!\langle\Psi_{b}\big|\Psi_{a}\rangle\!\rangle
=\delta_{ba}\Delta_{a}
\label{3.22}
\end{equation}
obeyed by the eigenfunctions to the problem we study here.

Equations (\ref{3.21}) and (\ref{3.17}) are important from the
theoretical point of view. However, it turns out that except for the
simplest case of a particle bound in the field of a single zero-range
potential (cf.\ Sec.\ \ref{VI.1}), this pair cannot be used to
practically perform the eigenfunction normalization process. The
reason for this is that if more than one potential center is involved,
the volume integral $\big<\Psi_{a}\big|
\Psi_{a}\big>_{\mathbb{R}_{\rho}^{3}}$ appearing in Eq.\ (\ref{3.17})
is not amenable to direct analytical evaluation. To overcome the
difficulty, we shall transform the formal definition (\ref{3.16}) of
the pseudo-product $\langle\!\langle\Psi_{b}\big|
\Psi_{a}\rangle\!\rangle$ to an operational form. To this end,
consider the easily provable operator identity
\begin{equation}
\mathcal{H}\beta^{(-)}-\beta^{(+)}\mathcal{H}+mc^{2}\mathcal{I}=0,
\label{3.23}
\end{equation}
which is obeyed by the Dirac Hamiltonian (\ref{3.4}). Premultiplying
Eq.\ (\ref{3.23}) with $\Psi_{b}^{\dag}(\boldsymbol{r})$,
postmultiplying with $\Psi_{a}(\boldsymbol{r})$ and integrating over
the domain $\mathbb{R}_{\rho}^{3}$ yields
\begin{equation}
\big<\Psi_{b}\big|
\mathcal{H}\beta^{(-)}\Psi_{a}\big>_{\mathbb{R}_{\rho}^{3}}
-\big<\Psi_{b}\big|
\beta^{(+)}\mathcal{H}\Psi_{a}\big>_{\mathbb{R}_{\rho}^{3}}
+mc^{2}\big<\Psi_{b}\big|\Psi_{a}\big>_{\mathbb{R}_{\rho}^{3}}=0.
\label{3.24}
\end{equation}
If in the first term on the left-hand side the action of $\mathcal{H}$
is transferred to the left with the aid of the integration by parts,
this gives
\begin{align}
& \big<\mathcal{H}\Psi_{b}\big|
\beta^{(-)}\Psi_{a}\big>_{\mathbb{R}_{\rho}^{3}}
+\frac{c\hbar}{\rho}\sum_{n=1}^{N}\big(\Psi_{b}\big|
\mathrm{i}\boldsymbol{\rho}_{n}\cdot\boldsymbol{\alpha}
\beta^{(-)}\Psi_{a}\big)_{\mathcal{S}_{n}}-\big<\Psi_{b}\big|
\beta^{(+)}\mathcal{H}\Psi_{a}\big>_{\mathbb{R}_{\rho}^{3}}
+mc^{2}\big<\Psi_{b}\big|\Psi_{a}\big>_{\mathbb{R}_{\rho}^{3}}=0,
\label{3.25}
\end{align}
where we have exploited the fact that the surface integral over the
infinite sphere $\mathcal{S}_{\infty}$ vanishes. Since
$\Psi_{a}(\boldsymbol{r})$ and $\Psi_{b}(\boldsymbol{r})$ are
eigenfunctions belonging to the energy eigenvalues $E_{a}$ and
$E_{b}$, respectively, and since the identity (\ref{2.8a}) holds, Eq.\
(\ref{3.25}) may be rewritten in the form
\begin{equation}
\big<\Psi_{b}\big|\big[E_{b}\beta^{(-)}-E_{a}\beta^{(+)}
+mc^{2}\mathcal{I}\big]\Psi_{a}\big>_{\mathbb{R}_{\rho}^{3}}
+\frac{c\hbar}{\rho}\sum_{n=1}^{N}
\big(\Psi_{b}\big|\mathrm{i}\boldsymbol{\rho}_{n}
\cdot\boldsymbol{\alpha}^{(+)}\Psi_{a}\big)_{\mathcal{S}_{n}}=0.
\label{3.26}
\end{equation}
In the limit $\rho\to0$, with the use of the interaction conditions
(\ref{2.7}), Eq.\ (\ref{3.26}) goes over into
\begin{align}
& \hspace*{-5em}
\lim_{\rho\to0}\left\{\big<\Psi_{b}\big|
\big[E_{b}\beta^{(-)}-E_{a}\beta^{(+)}+mc^{2}\mathcal{I}\big]
\Psi_{a}\big>_{\mathbb{R}_{\rho}^{3}}
-\frac{c\hbar\varepsilon_{a}k_{a}^{-1}}{\rho}
\sum_{n=1}^{N}\big(\Psi_{b}\big|
\beta^{(+)}\Psi_{a}\big)_{\mathcal{S}_{n}}\right\}
\nonumber \\
& =\frac{\hbar^{2}}{2m}\lim_{\rho\to0}
\sum_{n=1}^{N}\big(\Psi_{b}\big|\mathcal{K}_{n}^{(+)}
\Psi_{a}\big)_{\mathcal{S}_{n}}.
\label{3.27}
\end{align}
Using elementary properties of the matrices $\beta^{(\pm)}$, the above
result may be further transformed into the integral identity
\begin{align}
& \hspace*{-5em}
(E_{b}+mc^{2})\lim_{\rho\to0}
\left\{\big<\Psi_{b}\big|\Psi_{a}\big>_{\mathbb{R}_{\rho}^{3}}
-\frac{(\varepsilon_{b}k_{b}^{-1})
(\varepsilon_{a}k_{a}^{-1})}{\rho}
\sum_{n=1}^{N}\big(\Psi_{b}\big|
\beta^{(+)}\Psi_{a}\big)_{\mathcal{S}_{n}}\right\}
\nonumber \\
& =(E_{b}+E_{a})
\big<\Psi_{b}\big|\beta^{(+)}\Psi_{a}\big>_{\mathbb{R}^{3}}
+\frac{\hbar^{2}}{2m}\lim_{\rho\to0}
\sum_{n=1}^{N}\big(\Psi_{b}\big|\mathcal{K}_{n}^{(+)}
\Psi_{a}\big)_{\mathcal{S}_{n}},
\label{3.28}
\end{align}
where we have made use of the fact that
\begin{equation}
\lim_{\rho\to0}
\big<\Psi_{b}\big|\beta^{(+)}\Psi_{a}\big>_{\mathbb{R}_{\rho}^{3}}
=\big<\Psi_{b}\big|\beta^{(+)}\Psi_{a}\big>_{\mathbb{R}^{3}}.
\label{3.29}
\end{equation}
On invoking Eq.\ (\ref{3.16}), the left-hand side of Eq.\ (\ref{3.28})
is seen to be a multiple of the pseudo-product
$\langle\!\langle\Psi_{b}\big|\Psi_{a}\rangle\!\rangle$. If the
orthogonality relation (\ref{3.15}) holds, which we shall assume to be
the case, then irrespective of whether the eigenfunctions have been
normalized in the sense of Eq.\ (\ref{3.21}) or not, Eq.\ (\ref{3.28})
may be cast into the symmetric form
\begin{equation}
\langle\!\langle\Psi_{b}\big|\Psi_{a}\rangle\!\rangle
=\frac{\displaystyle(E_{b}+E_{a})
\big<\Psi_{b}\big|\beta^{(+)}\Psi_{a}\big>_{\mathbb{R}^{3}}
+\frac{\hbar^{2}}{2m}\lim_{\rho\to0}
\sum_{n=1}^{N}\big(\Psi_{b}\big|\mathcal{K}_{n}^{(+)}
\Psi_{a}\big)_{\mathcal{S}_{n}}}{\sqrt{(E_{b}+mc^{2})(E_{a}+mc^{2})}}.
\label{3.30}
\end{equation}
This is the sought alternative representation for
$\langle\!\langle\Psi_{b}\big|\Psi_{a}\rangle\!\rangle$. In the
particular case of $b=a$, it becomes
\begin{equation}
\langle\!\langle\Psi_{a}\big|\Psi_{a}\rangle\!\rangle
=\big(1-\varepsilon_{a}^{2}\big)
\big<\Psi_{a}\big|\beta^{(+)}\Psi_{a}\big>_{\mathbb{R}^{3}}
+\frac{\hbar}{2mc}\varepsilon_{a}k_{a}^{-1}\lim_{\rho\to0}
\sum_{n=1}^{N}\big(\Psi_{a}\big|\mathcal{K}_{n}^{(+)}
\Psi_{a}\big)_{\mathcal{S}_{n}}.
\label{3.31}
\end{equation}

The practical advantage of the representation of
$\langle\!\langle\Psi_{a}\big|\Psi_{a}\rangle\!\rangle$ given in Eq.\
(\ref{3.31}) over the one in Eq.\ (\ref{3.17}) is that the surface and
the volume integrals appearing in the former may be reduced
analytically to closed-form algebraic expressions. Using Eqs.\
(\ref{2.2b}) and (\ref{2.3a}), with no difficulty one shows that
\begin{equation}
\lim_{\rho\to0}\big(\Psi_{a}\big|
\mathcal{K}_{n}^{(+)}\Psi_{a}\big)_{\mathcal{S}_{n}}
=\frac{4\pi}{k_{a}^{2}}\chi_{an}^{\dag}K_{n}\chi_{an}.
\label{3.32}
\end{equation}
Reduction of the volume integral
$\big<\Psi_{a}\big|\beta^{(+)}\Psi_{a}\big>_{\mathbb{R}^{3}}$ is a bit
more involved. We have
\begin{equation}
\big<\Psi_{a}\big|\beta^{(+)}\Psi_{a}\big>_{\mathbb{R}^{3}}
=\frac{1}{k_{a}^{2}}\sum_{n,n'=1}^{N}\chi_{an}^{\dag}\chi_{an'}
\int_{\mathbb{R}^{3}}\mathrm{d}^{3}\boldsymbol{r}\:
\frac{\mathrm{e}^{-k_{a}|\boldsymbol{r}-\boldsymbol{r}_{n}|}}
{|\boldsymbol{r}-\boldsymbol{r}_{n}|}
\frac{\mathrm{e}^{-k_{a}|\boldsymbol{r}-\boldsymbol{r}_{n'}|}}
{|\boldsymbol{r}-\boldsymbol{r}_{n'}|}.
\label{3.33}
\end{equation}
The integral in Eq.\ (\ref{3.33}) may be evaluated in the prolate
spheroidal coordinates $\xi_{nn'}$, $\eta_{nn'}$, $\varphi_{nn'}$. The
former two are defined as
\begin{subequations}
\begin{equation}
\xi_{nn'}=\frac{|\boldsymbol{r}-\boldsymbol{r}_{n}|
+|\boldsymbol{r}-\boldsymbol{r}_{n'}|}
{|\boldsymbol{r}_{n}-\boldsymbol{r}_{n'}|}
\label{3.34a}
\end{equation}
and
\begin{equation}
\eta_{nn'}=\frac{|\boldsymbol{r}-\boldsymbol{r}_{n}|
-|\boldsymbol{r}-\boldsymbol{r}_{n'}|}
{|\boldsymbol{r}_{n}-\boldsymbol{r}_{n'}|},
\label{3.34b}
\end{equation}
\label{3.34}%
\end{subequations}
respectively, whereas the latter is a rotational angle in a plane
perpendicular to the vector $\boldsymbol{r}_{n}-\boldsymbol{r}_{n'}$.
The ranges in which these coordinates vary are
\begin{equation}
1\leqslant\xi_{nn'}<\infty,
\qquad
-1\leqslant\eta_{nn'}\leqslant1,
\qquad
0\leqslant\varphi_{nn'}<2\pi.
\label{3.35}
\end{equation}
Since from Eqs.\ (\ref{3.34a}) and (\ref{3.34b}) one has
\begin{subequations}
\begin{equation}
|\boldsymbol{r}-\boldsymbol{r}_{n}|
=\frac{1}{2}|\boldsymbol{r}_{n}-\boldsymbol{r}_{n'}|
(\xi_{nn'}+\eta_{nn'}),
\label{3.36a}
\end{equation}
\begin{equation}
|\boldsymbol{r}-\boldsymbol{r}_{n'}|
=\frac{1}{2}|\boldsymbol{r}_{n}-\boldsymbol{r}_{n'}|
(\xi_{nn'}-\eta_{nn'}),
\label{3.36b}
\end{equation}
\label{3.36}%
\end{subequations}
and since in the prolate spheroidal coordinates the infinitesimal
volume element $\mathrm{d}^{3}\boldsymbol{r}$ is
\begin{equation}
\mathrm{d}^{3}\boldsymbol{r}
=\frac{|\boldsymbol{r}_{n}-\boldsymbol{r}_{n'}|^{3}}{8}
\left(\xi_{nn'}^{2}-\eta_{nn'}^{2}\right)
\mathrm{d}\xi_{nn'}\mathrm{d}\eta_{nn'}\mathrm{d}\varphi_{nn'},
\label{3.37}
\end{equation}
elementary integrations over the three variables yield the result
\begin{equation}
\int_{\mathbb{R}^{3}}\mathrm{d}^{3}\boldsymbol{r}\:
\frac{\mathrm{e}^{-k_{a}|\boldsymbol{r}-\boldsymbol{r}_{n}|}}
{|\boldsymbol{r}-\boldsymbol{r}_{n}|}
\frac{\mathrm{e}^{-k_{a}|\boldsymbol{r}-\boldsymbol{r}_{n'}|}}
{|\boldsymbol{r}-\boldsymbol{r}_{n'}|}
=\frac{2\pi}{k_{a}}
\mathrm{e}^{-k_{a}|\boldsymbol{r}_{n}-\boldsymbol{r}_{n'}|}.
\label{3.38}
\end{equation}
It then follows that
\begin{equation}
\big<\Psi_{a}\big|\beta^{(+)}\Psi_{a}\big>_{\mathbb{R}^{3}}
=\frac{2\pi}{k_{a}^{3}}\sum_{n,n'=1}^{N}\chi_{an}^{\dag}\chi_{an'}
\mathrm{e}^{-k_{a}|\boldsymbol{r}_{n}-\boldsymbol{r}_{n'}|}.
\label{3.39}
\end{equation}
On combining Eqs.\ (\ref{3.31}), (\ref{3.32}) and (\ref{3.39}), we
infer the following algebraic representation of the pseudo-product
$\langle\!\langle\Psi_{a}\big|\Psi_{a}\rangle\!\rangle$:
\begin{equation}
\langle\!\langle\Psi_{a}\big|\Psi_{a}\rangle\!\rangle
=\frac{2\pi}{k_{a}^{3}}\Bigg[\big(1-\varepsilon_{a}^{2}\big)
\sum_{n,n'=1}^{N}\chi_{an}^{\dag}\chi_{an'}
\mathrm{e}^{-k_{a}|\boldsymbol{r}_{n}-\boldsymbol{r}_{n'}|}
+\frac{\hbar\varepsilon_{a}}{mc}
\sum_{n=1}^{N}\chi_{an}^{\dag}K_{n}\chi_{an}\Bigg].
\label{3.40}
\end{equation}
Hence, the relation (\ref{3.21}) may be rewritten in the following
form:
\begin{equation}
\frac{2\pi}{k_{a}^{3}}\Bigg[\big(1-\varepsilon_{a}^{2}\big)
\sum_{n,n'=1}^{N}\chi_{an}^{\dag}\chi_{an'}
\mathrm{e}^{-k_{a}|\boldsymbol{r}_{n}-\boldsymbol{r}_{n'}|}
+\frac{\hbar\varepsilon_{a}}{mc}
\sum_{n=1}^{N}\chi_{an}^{\dag}K_{n}\chi_{an}\Bigg]=\Delta_{a}.
\label{3.41}
\end{equation}
If $\Delta_{a}\neq0$, Eq.\ (\ref{3.41}) fixes the absolute value of a
common multiplicative factor in the spinors $\chi_{an}$, and thus
actually normalizes $\Psi_{a}(\boldsymbol{r})$.

It is profitable to juggle a bit with the form of Eq.\ (\ref{3.40}).
If we transform its right-hand side with the aid of Eq.\ (\ref{2.4b}),
this gives
\begin{equation}
\langle\!\langle\Psi_{a}\big|\Psi_{a}\rangle\!\rangle
=\frac{4\pi\varepsilon_{a}}{c\hbar k_{a}^{3}}
\Bigg[E_{a}k_{a}^{-1}\sum_{n,n'=1}^{N}\chi_{an}^{\dag}\chi_{an'}
\mathrm{e}^{-k_{a}|\boldsymbol{r}_{n}-\boldsymbol{r}_{n'}|}
+\frac{\hbar^{2}}{2m}
\sum_{n=1}^{N}\chi_{an}^{\dag}K_{n}\chi_{an}\Bigg].
\label{3.42}
\end{equation}
It is immediately seen that the expression in the square bracket
coincides with the left-hand side of Eq.\ (\ref{2.20}). Hence, we get
the remarkable relationship
\begin{equation}
\langle\!\langle\Psi_{a}\big|\Psi_{a}\rangle\!\rangle
=4\pi c\hbar\varepsilon_{a}k_{a}^{-2}\mathsf{x}_{a}^{\dag}
\left[\frac{\partial\mathsf{L}(E)}{\partial E}\right]_{E=E_{a}}
\mathsf{x}_{a}.
\label{3.43}
\end{equation}
If it is combined with the normalization relation (\ref{3.21}), the
latter takes the form
\begin{equation}
4\pi c\hbar\varepsilon_{a}k_{a}^{-2}\mathsf{x}_{a}^{\dag}
\left[\frac{\partial\mathsf{L}(E)}{\partial E}\right]_{E=E_{a}}
\mathsf{x}_{a}=\Delta_{a}.
\label{3.44}
\end{equation}
Equation (\ref{3.44}) will find an application at the end of the next
section, where we shall exploit it to derive a useful relationship
between a normalized eigenfunction $\Psi_{a}(\boldsymbol{r})$ and a
related Sturmian function $\Sigma_{a}(E,\boldsymbol{r})$.
%
%\newpage
%
\section{The Sturmian functions}
\label{IV}
\setcounter{equation}{0}
The Sturmian functions for our model are defined as these solutions to
the Dirac equation
\begin{equation}
[-\mathrm{i}c\hbar\boldsymbol{\alpha}\cdot\boldsymbol{\nabla}
+mc^{2}\beta-E\mathcal{I}]\Sigma_{a}(E,\boldsymbol{r})=0
\qquad (\mbox{$\boldsymbol{r}\neq\boldsymbol{r}_{n}$; 
$n=1,\ldots,N$}),
\label{4.1}
\end{equation}
which are of the form
\begin{equation}
\Sigma_{a}(E,\boldsymbol{r})
=\sum_{n=1}^{N}
\left(
\begin{array}{c}
f(k|\boldsymbol{r}-\boldsymbol{r}_{n}|)\eta_{an}(E) \\*[1ex]
\mathrm{i}\varepsilon g(k|\boldsymbol{r}-\boldsymbol{r}_{n}|)
\boldsymbol{\mu}_{n}(\boldsymbol{r})
\cdot\boldsymbol{\sigma}\eta_{an}(E)
\end{array}
\right),
\label{4.2}
\end{equation}
with $f(z)$, $g(z)$, $\mu_{n}(\boldsymbol{r})$, $k$ and $\varepsilon$
defined in Eqs.\ (\ref{2.3}), (\ref{2.5}) and (\ref{2.16}),
respectively, and which are forced to obey the constraining conditions
\begin{align}
& \lim_{\boldsymbol{r}\to\boldsymbol{r}_{n}}\left\{\mathrm{i}
(\boldsymbol{r}-\boldsymbol{r}_{n})\cdot\boldsymbol{\alpha}^{(+)}
+\frac{\hbar}{2mc}|\boldsymbol{r}-\boldsymbol{r}_{n}|
\mathcal{K}_{n}^{(+)}
-\lambda_{a}(E)\varepsilon
|\boldsymbol{r}-\boldsymbol{r}_{n}|\beta^{(+)}
+\varepsilon k^{-1}\beta^{(+)}\right\}\Sigma_{a}(E,\boldsymbol{r})=0
\nonumber \\*
& \hspace*{30em} (n=1,\ldots,N)
\label{4.3}
\end{align}
[cf.\ Eq.\ (\ref{2.7})], with $\mathcal{K}_{n}^{(+)}$ defined as in
Eqs.\ (\ref{2.9})--(\ref{2.11}). In Eqs.\ (\ref{4.1})--(\ref{4.3}),
$E$ is presumed to have a \emph{fixed\/} value from the range
$-mc^{2}<E\leqslant mc^{2}$ [in general, $E$ need not coincide with
any of the eigenenergies to the eigenproblem constituted by Eqs.\
(\ref{2.1}), (\ref{2.2}) and (\ref{2.7})], whereas the role of an
eigenparameter is now taken over by the parameter $\lambda_{a}(E)$
entering the limiting conditions (\ref{4.3}). The two-component
spinors $\eta_{an}(E)$ [not to be confused with the spheroidal
coordinate $\eta_{nn'}$ defined in Eq.\ (\ref{3.34b})] entering Eq.\
(\ref{4.2}) play the role of generalized linear combination
coefficients and may be determined by solving the algebraic
eigensystem
\begin{equation}
\left\{\frac{\hbar}{2mc\varepsilon}K_{n}
-[\lambda_{a}(E)+1]I\right\}\eta_{an}(E)
+\sum_{\substack{n'=1 \\ (n'\neq n)}}^{N}
f(k|\boldsymbol{r}_{n}-\boldsymbol{r}_{n'}|)\eta_{an'}(E)=0
\qquad (n=1,\ldots,N)
\label{4.4}
\end{equation}
emerging after Eq.\ (\ref{4.2}) is inserted into Eq.\ (\ref{4.3}). If
the spinors $\eta_{an}(E)$ are collected in a $2N$-component vector
\begin{equation}
\mathsf{y}_{a}(E)
=\left(
\begin{array}{ccc}
\eta_{a1}^\mathrm{T}(E) & \cdots & \eta_{aN}^\mathrm{T}(E)
\end{array}
\right)^{\mathrm{T}},
\label{4.5} 
\end{equation}
the eigensystem (\ref{4.4}) may be rewritten compactly as
\begin{equation}
\mathsf{L}(E)\mathsf{y}_{a}(E)=\lambda_{a}(E)\mathsf{y}_{a}(E),
\label{4.6}
\end{equation}
where $\mathsf{L}(E)$ is the $2N\times2N$ matrix with its $2\times2$
block-elements defined in Eq.\ (\ref{2.15}). We see that
$\lambda_{a}(E)$ and $\mathsf{y}_{a}(E)$ are an eigenvalue and an
associated eigenvector of the matrix $\mathsf{L}(E)$, respectively.
Since $k$ and $\varepsilon$ are real, $\mathsf{L}(E)$ is Hermitian and
therefore we know in advance that all its eigenvalues $\lambda_{a}(E)$
are real, and also that eigenvectors belonging to different
eigenvalues are orthogonal in the sense of
\begin{equation}
\mathsf{y}_{b}^{\dag}(E)\mathsf{y}_{a}(E)=0
\qquad [\lambda_{b}(E)\neq\lambda_{a}(E)].
\label{4.7}
\end{equation}
In what follows, we shall be assuming that eigenvectors associated
with degenerate eigenvalues (if there are any) have been
orthogonalized in the same manner, and consequently it holds that
\begin{equation}
\mathsf{y}_{b}^{\dag}(E)\mathsf{y}_{a}(E)=0
\qquad (b\neq a).
\label{4.8}
\end{equation}
Temporarily, we leave aside the issue of normalization of the
eigenvectors $\mathsf{y}_{a}(E)$ and turn to the problem of
orthogonality and normalization of the Sturmian functions.

To this end, consider the volume integral $\big<\Sigma_{b}(E)\big|
[\mathcal{H}-E\mathcal{I}]\Sigma_{a}(E)\big>_{\mathbb{R}_{\rho}^{3}}$
over the domain $\mathbb{R}_{\rho}^{3}$ defined in Eq.\ (\ref{3.1}).
If the action of the operator $\mathcal{H}-E\mathcal{I}$ is
transformed to the left with the use of the Gauss integral formula,
this leads to the identity
\begin{equation}
\big<\Sigma_{b}(E)\big|
[\mathcal{H}-E\mathcal{I}]\Sigma_{a}(E)\big>_{\mathbb{R}_{\rho}^{3}}
=\big<[\mathcal{H}-E\mathcal{I}]\Sigma_{b}(E)\big|
\Sigma_{a}(E)\big>_{\mathbb{R}_{\rho}^{3}}
+\frac{c\hbar}{\rho}\sum_{n=1}^{N}\big(\Sigma_{b}(E)\big|
\mathrm{i}\boldsymbol{\rho}_{n}\cdot\boldsymbol{\alpha}
\Sigma_{a}(E)\big)_{\mathcal{S}_{n}},
\label{4.9}
\end{equation}
where the integral over the surface of an infinitely distant sphere
$\mathcal{S}_{\infty}$ has been omitted, being zero in view of the
exponential decay of both $\Sigma_{a}(E,\boldsymbol{r})$ and
$\Sigma_{b}(E,\boldsymbol{r})$. Since the operator
$\mathcal{H}-E\mathcal{I}$ annihilates both Sturmians
$\Sigma_{a}(E,\boldsymbol{r})$ and $\Sigma_{b}(E,\boldsymbol{r})$
[cf.\ Eq.\ (\ref{4.1})], the two volume integrals in Eq.\ (\ref{4.9})
vanish, yielding
\begin{equation}
\frac{1}{\rho}\sum_{n=1}^{N}\big(\Sigma_{b}(E)\big|
\mathrm{i}\boldsymbol{\rho}_{n}\cdot\boldsymbol{\alpha}
\Sigma_{a}(E)\big)_{\mathcal{S}_{n}}=0
\label{4.10}
\end{equation}
and then, with the aid of Eqs.\ (\ref{3.6}) and (\ref{3.7}),
\begin{equation}
\frac{1}{\rho}\sum_{n=1}^{N}\big(\Sigma_{b}(E)\big|
\mathrm{i}\boldsymbol{\rho}_{n}\cdot\boldsymbol{\alpha}^{(+)}
\Sigma_{a}(E)\big)_{\mathcal{S}_{n}}
-\frac{1}{\rho}\sum_{n=1}^{N}
\big(\mathrm{i}\boldsymbol{\rho}_{n}\cdot\boldsymbol{\alpha}^{(+)}
\Sigma_{b}(E)\big|\Sigma_{a}(E)\big)_{\mathcal{S}_{n}}=0.
\label{4.11}
\end{equation}
If we let the common radius $\rho$ of the spheres $\mathcal{S}_{n}$
tend to zero, after exploiting the constraints (\ref{4.3}) we obtain
\begin{equation}
[\lambda_{a}(E)-\lambda_{b}(E)]
\lim_{\rho\to0}\sum_{n=1}^{N}\big(\Sigma_{b}(E)\big|
\beta^{(+)}\Sigma_{a}(E)\big)_{\mathcal{S}_{n}}=0,
\label{4.12}
\end{equation}
where we have also made use of the fact that the Sturmian eigenvalues
are real [cf.\ the remark preceding Eq.\ (\ref{4.7})]. Equation
(\ref{4.12}) implies that the Sturmian functions obey the
orthogonality relation
\begin{equation}
\lim_{\rho\to0}\sum_{n=1}^{N}\big(\Sigma_{b}(E)\big|
\beta^{(+)}\Sigma_{a}(E)\big)_{\mathcal{S}_{n}}=0
\qquad [\lambda_{b}(E)\neq\lambda_{a}(E)].
\label{4.13}
\end{equation}
Actually, Eq.\ (\ref{4.13}) does not offer anything more than Eq.\
(\ref{4.7}) does. Indeed, using Eq.\ (\ref{4.2}) it is straightforward
to show that
\begin{equation}
\lim_{\rho\to0}\big(\Sigma_{b}(E)\big|
\beta^{(+)}\Sigma_{a}(E)\big)_{\mathcal{S}_{n}}
=\frac{4\pi}{k^{2}}\eta_{bn}^{\dag}(E)\eta_{an}(E),
\label{4.14}
\end{equation}
hence, it follows that
\begin{equation}
\lim_{\rho\to0}\sum_{n=1}^{N}\big(\Sigma_{b}(E)\big|
\beta^{(+)}\Sigma_{a}(E)\big)_{\mathcal{S}_{n}}
=\frac{4\pi}{k^{2}}\mathsf{y}_{b}^{\dag}(E)\mathsf{y}_{a}(E),
\label{4.15}
\end{equation}
which implies the equivalence of Eqs.\ (\ref{4.13}) and (\ref{4.7}).
But if Eq.\ (\ref{4.15}) is combined with Eq.\ (\ref{4.8}), one
obtains a still more general orthogonality relation, namely
\begin{equation}
\lim_{\rho\to0}\sum_{n=1}^{N}\big(\Sigma_{b}(E)\big|
\beta^{(+)}\Sigma_{a}(E)\big)_{\mathcal{S}_{n}}=0
\qquad (b\neq a).
\label{4.16}
\end{equation}
If we normalize the Sturmian functions in accordance with
\begin{equation}
\lim_{\rho\to0}\sum_{n=1}^{N}\big(\Sigma_{a}(E)\big|
\beta^{(+)}\Sigma_{a}(E)\big)_{\mathcal{S}_{n}}=1,
\label{4.17}
\end{equation}
we have the integral orthonormality relation
\begin{equation}
\lim_{\rho\to0}\sum_{n=1}^{N}\big(\Sigma_{b}(E)\big|
\beta^{(+)}\Sigma_{a}(E)\big)_{\mathcal{S}_{n}}=\delta_{ba}.
\label{4.18}
\end{equation}
Concluding this thread, we observe that imposing the constraint
(\ref{4.17}) we have automatically normalized the eigenvectors
$\mathsf{y}_{a}(E)$, so that Eq.\ (\ref{4.8}) may be replaced with the
more general algebraic orthonormality relation
\begin{equation}
\frac{4\pi}{k^{2}}\mathsf{y}_{b}^{\dag}(E)\mathsf{y}_{a}(E)
=\delta_{ba}.
\label{4.19}
\end{equation}

Until this moment, the index used to distinguish between different
Sturmian eigenpairs $\lambda_{a}(E)$ and
$\Sigma_{a}(E,\boldsymbol{r})$ has not been related in any way to the
index labeling particle's eigenenergies $E_{a}$ and their associated
eigenfunctions $\Psi_{a}(\boldsymbol{r})$. However, it is convenient
to correlate these indices to have
\begin{equation}
\lambda_{a}(E_{a})=0
\quad \textrm{and} \quad
\mathsf{y}_{a}(E_{a})=A_{a}\mathsf{x}_{a},
\label{4.20}
\end{equation}
where $A_{a}$ is a proportionality factor which is to be determined.
Then it holds that

\begin{equation}
\Sigma_{a}(E_{a},\boldsymbol{r})
=A_{a}\Psi_{a}(\boldsymbol{r}).
\label{4.21}
\end{equation}
To determine the coefficient $A_{a}$, we invoke the Hellmann--Feynman
theorem for the matrix $\mathsf{L}(E)$, which is
\begin{equation}
\mathsf{y}_{a}^{\dag}(E)\frac{\partial\mathsf{L}(E)}{\partial E}
\mathsf{y}_{a}(E)=\frac{\partial\lambda_{a}(E)}{\partial E}
\mathsf{y}_{a}^{\dag}(E)\mathsf{y}_{a}(E).
\label{4.22}
\end{equation}
After the limit $E\to E_{a}$ is taken and then, on the left-hand side
only, the use is made of the second of Eqs.\ (\ref{4.20}), Eq.\
(\ref{4.22}) becomes
\begin{equation}
|A_{a}|^{2}\mathsf{x}_{a}^{\dag}
\left[\frac{\partial\mathsf{L}(E)}{\partial E}\right]_{E=E_{a}}
\mathsf{x}_{a}
=\left[\frac{\partial\lambda_{a}(E)}{\partial E}\right]_{E=E_{a}}
\mathsf{y}_{a}^{\dag}(E_{a})\mathsf{y}_{a}(E_{a}).
\label{4.23}
\end{equation}
A simplification occurs after the left-hand side of Eq.\ (\ref{4.23})
is transformed with the aid of Eq.\ (\ref{3.44}) and the right-hand
side with the aid of Eq.\ (\ref{4.19}), the latter being taken in the
case of $b=a$ and $E=E_{a}$. This gives
\begin{equation}
|A_{a}|^{2}\Delta_{a}=c\hbar\varepsilon_{a}
\left[\frac{\partial\lambda_{a}(E)}{\partial E}\right]_{E=E_{a}}.
\label{4.24}
\end{equation}
From this one finds that
\begin{equation}
A_{a}=\sqrt{c\hbar\varepsilon_{a}
|\partial\lambda_{a}(E)/\partial E|_{E=E_{a}}}
\qquad (\Delta_{a}\neq0),
\label{4.25}
\end{equation}
an adjustable phase factor in $A_{a}$ being chosen to have $A_{a}$
real and positive, and also that
\begin{equation}
\Delta_{a}
=\sgn\left[\frac{\partial\lambda_{a}(E)}{\partial E}\right]_{E=E_{a}}.
\label{4.26}
\end{equation}

In summary, we see that once the Sturmian eigenpairs $\lambda_{a}(E)$
and $\Sigma_{a}(E,\boldsymbol{r})$ have been found, with
$\Sigma_{a}(E,\boldsymbol{r})$ normalized in the sense of Eq.\
(\ref{4.17}), one may determine the particle's eigenenergies $E_{a}$
from the first of Eqs.\ (\ref{4.20}), whereas the associated
eigenfunctions $\Psi_{a}(\boldsymbol{r})$, normalized in the sense of
Eq.\ (\ref{3.21}), are given by
\begin{equation}
\Psi_{a}(\boldsymbol{r})
=\frac{\Sigma_{a}(E_{a},\boldsymbol{r})}{\sqrt{c\hbar\varepsilon_{a}
|\partial\lambda_{a}(E)/\partial E|_{E=E_{a}}}}
\qquad (\Delta_{a}\neq0).
\label{4.27}
\end{equation}
It should be observed that if $E_{a}$ is degenerate, the associated
eigenfunctions resulting from Eq.\ (\ref{4.27}) may need to be
orthogonalized to obey the orthonormality relation (\ref{3.22}).
%
%\newpage
%
\section{The matrix Green's function and its Sturmian representation}
\label{V}
\setcounter{equation}{0}
In our model, the matrix Green's function
$\mathcal{G}(E,\boldsymbol{r},\boldsymbol{r}')$ due to a source
located at the point $\boldsymbol{r}'\neq\boldsymbol{r}_{n}$,
$n=1,\ldots,N$, satisfies the inhomogeneous differential equation
\begin{equation}
[-\mathrm{i}c\hbar\boldsymbol{\alpha}\cdot\boldsymbol{\nabla}
+mc^{2}\beta-E\mathcal{I}]
\mathcal{G}(E,\boldsymbol{r},\boldsymbol{r}')
=\delta^{(3)}(\boldsymbol{r}-\boldsymbol{r}')\mathcal{I}
\qquad (\mbox{$\boldsymbol{r},\boldsymbol{r}'\neq\boldsymbol{r}_{n}$; 
$n=1,\ldots,N$}),
\label{5.1}
\end{equation}
the asymptotic condition
\begin{equation}
\mathcal{G}(E,\boldsymbol{r},\boldsymbol{r}')
\stackrel{r\to\infty}{\longrightarrow}\mathcal{A}(E,\boldsymbol{r}')
\frac{\mathrm{e}^{-kr}}{r},
\label{5.2}
\end{equation}
where $\mathcal{A}(E,\boldsymbol{r}')$ is a certain $4\times4$
amplitude matrix, and also the limiting constraints
\begin{equation}
\lim_{\boldsymbol{r}\to\boldsymbol{r}_{n}}
\left[\mathrm{i}(\boldsymbol{r}-\boldsymbol{r}_{n})
\cdot\boldsymbol{\alpha}^{(+)}
+\frac{\hbar}{2mc}|\boldsymbol{r}-\boldsymbol{r}_{n}|
\mathcal{K}_{n}^{(+)}+\varepsilon k^{-1}\beta^{(+)}\right]
\mathcal{G}(E,\boldsymbol{r},\boldsymbol{r}')=0
\qquad (n=1,\ldots,N)
\label{5.3}
\end{equation}
at locations of the potential centers [cf.\ Eq.\ (\ref{2.7})]. The
energy parameter $E$ is constrained to the interval
\mbox{$-mc^{2}<E\leqslant mc^{2}$}. We shall seek
$\mathcal{G}(E,\boldsymbol{r},\boldsymbol{r}')$ in the form
\begin{equation}
\mathcal{G}(E,\boldsymbol{r},\boldsymbol{r}')
=\mathcal{G}_{0}(E,\boldsymbol{r},\boldsymbol{r}')
+\sum_{a=1}^{2N}\Sigma_{a}(E,\boldsymbol{r})
C_{a}^{\dag}(E,\boldsymbol{r}'),
\label{5.4}
\end{equation}
where
\begin{align}
\mathcal{G}_{0}(E,\boldsymbol{r},\boldsymbol{r}') 
& =\frac{1}{4\pi c^{2}\hbar^{2}}
[-\mathrm{i}c\hbar\boldsymbol{\alpha}\cdot\boldsymbol{\nabla}
+mc^{2}\beta+E\mathcal{I}]
\frac{\mathrm{e}^{-k|\boldsymbol{r}-\boldsymbol{r}'|}}
{|\boldsymbol{r}-\boldsymbol{r}'|}
\nonumber \\
& =\frac{k}{4\pi c^{2}\hbar^{2}}
\left(
\begin{array}{cc}
(E+mc^{2})f(k|\boldsymbol{r}-\boldsymbol{r}'|)I &
\mathrm{i}c\hbar kg(k|\boldsymbol{r}-\boldsymbol{r}'|)
\boldsymbol{\mu}(\boldsymbol{r},\boldsymbol{r}')
\cdot\boldsymbol{\sigma}
\\*[1ex]
\mathrm{i}c\hbar kg(k|\boldsymbol{r}-\boldsymbol{r}'|)
\boldsymbol{\mu}(\boldsymbol{r},\boldsymbol{r}')
\cdot\boldsymbol{\sigma} &
(E-mc^{2})f(k|\boldsymbol{r}-\boldsymbol{r}'|)I
\end{array}
\right),
\label{5.5}
\end{align}
with
\begin{equation}
\boldsymbol{\mu}(\boldsymbol{r},\boldsymbol{r}')
=\frac{\boldsymbol{r}-\boldsymbol{r}'}
{|\boldsymbol{r}-\boldsymbol{r}'|},
\label{5.6}
\end{equation}
is the free-particle Dirac--Green's function,
$\Sigma_{a}(E,\boldsymbol{r})$ are the Sturmian functions of Sec.\
\ref{IV}, while $C_{a}^{\dag}(E,\boldsymbol{r})$ are spinor expansion 
coefficients which remain to be determined. From the fact that
$\mathcal{G}_{0}(E,\boldsymbol{r},\boldsymbol{r}')$ is known to obey
the inhomogeneous equation
\begin{equation}
[-\mathrm{i}c\hbar\boldsymbol{\alpha}\cdot\boldsymbol{\nabla}
+mc^{2}\beta-E\mathcal{I}]
\mathcal{G}_{0}(E,\boldsymbol{r},\boldsymbol{r}')
=\delta^{(3)}(\boldsymbol{r}-\boldsymbol{r}')\mathcal{I},
\label{5.7}
\end{equation}
whereas the Sturmian functions solve the homogeneous equation
(\ref{4.1}), we see that the function
$\mathcal{G}(E,\boldsymbol{r},\boldsymbol{r}')$ defined above does
indeed satisfy the inhomogeneous equation (\ref{5.1}). In turn, it
follows from Eqs.\ (\ref{5.5}), (\ref{4.2}) and (\ref{2.3}) that the
asymptotic condition (\ref{5.2}) is also fulfilled. Hence, it remains
to adjust the coefficients $C_{a}^{\dag}(E,\boldsymbol{r}')$ so that
the constraints (\ref{5.3}) are complied with.

To achieve the above goal, at first we observe that if the source
point $\boldsymbol{r}'$ is located in the domain
$\mathbb{R}_{\rho}^{3}$ defined in Eq.\ (\ref{3.1}), from Eqs.\
(\ref{5.1}) one has
\begin{equation}
\big<\Sigma_{a}(E)\big|[\mathcal{H}-E\mathcal{I}]
\mathcal{G}(E,\boldsymbol{r}')\big>_{\mathbb{R}_{\rho}^{3}}
=\Sigma_{a}^{\dag}(E,\boldsymbol{r}'),
\label{5.8}
\end{equation}
where $\mathcal{H}$ stands for the Dirac Hamiltonian (\ref{3.4}). Then
it trivially follows that
\begin{equation}
\lim_{\rho\to0}\big<\Sigma_{a}(E)\big|[\mathcal{H}-E\mathcal{I}]
\mathcal{G}(E,\boldsymbol{r}')\big>_{\mathbb{R}_{\rho}^{3}}
=\Sigma_{a}^{\dag}(E,\boldsymbol{r}').
\label{5.9}
\end{equation}
On the other hand, if in the integral $\big<\Sigma_{a}(E)\big|
[\mathcal{H}-E\mathcal{I}]
\mathcal{G}(E,\boldsymbol{r}')\big>_{\mathbb{R}_{\rho}^{3}}$ action of
the operator $\mathcal{H}-E\mathcal{I}$ is transferred to the left,
with the aid of the Gauss divergence formula one obtains
\begin{align}
\big<\Sigma_{a}(E)\big|[\mathcal{H}-E\mathcal{I}]
\mathcal{G}(E,\boldsymbol{r}')\big>_{\mathbb{R}_{\rho}^{3}}
& =\big<[\mathcal{H}-E\mathcal{I}]\Sigma_{a}(E)\big|
\mathcal{G}(E,\boldsymbol{r}')\big>_{\mathbb{R}_{\rho}^{3}}
+\frac{c\hbar}{\rho}
\sum_{n=1}^{N}\big(\Sigma_{a}(E)\big|
\mathrm{i}\boldsymbol{\rho}_{n}\cdot\boldsymbol{\alpha}
\mathcal{G}(E,\boldsymbol{r}')\big)_{\mathcal{S}_{n}},
\label{5.10}
\end{align}
the omitted integral over the infinite sphere $\mathcal{S}_{\infty}$
being zero. Now, the volume integral on the right-hand side vanishes
by virtue of Eq.\ (\ref{4.1}), while use of Eqs.\ (\ref{3.6}) and
(\ref{3.7}) splits each of the surface integrals into two ones. This
leads to
\begin{align}
& \hspace*{-2em}
\big<\Sigma_{a}(E)\big|[\mathcal{H}-E\mathcal{I}]
\mathcal{G}(E,\boldsymbol{r}')\big>_{\mathbb{R}_{\rho}^{3}}
\nonumber \\*
& =\frac{c\hbar}{\rho}\sum_{n=1}^{N}\big(\Sigma_{a}(E)\big|
\mathrm{i}\boldsymbol{\rho}_{n}\cdot\boldsymbol{\alpha}^{(+)}
\mathcal{G}(E,\boldsymbol{r}')\big)_{\mathcal{S}_{n}}
-\frac{c\hbar}{\rho}\sum_{n=1}^{N}
\big(\mathrm{i}\boldsymbol{\rho}_{n}\cdot\boldsymbol{\alpha}^{(+)}
\Sigma_{a}(E)\big|
\mathcal{G}(E,\boldsymbol{r}')\big)_{\mathcal{S}_{n}}.
\label{5.11}
\end{align}
Applying the limit $\rho\to0$ to both sides of Eq.\ (\ref{5.11}) and
transforming the right-hand side with the aid of the limiting
relations (\ref{4.3}) and (\ref{5.3}) gives
\begin{equation}
\lim_{\rho\to0}\big<\Sigma_{a}(E)\big|[\mathcal{H}-E\mathcal{I}]
\mathcal{G}(E,\boldsymbol{r}')\big>_{\mathbb{R}_{\rho}^{3}}
=-c\hbar\varepsilon\lambda_{a}(E)
\lim_{\rho\to0}\sum_{n=1}^{N}\big(\Sigma_{a}(E)\big|
\beta^{(+)}\mathcal{G}(E,\boldsymbol{r}')\big)_{\mathcal{S}_{n}}.
\label{5.12}
\end{equation}
Now, from Eq.\ (\ref{5.4}) one has
\begin{align}
& \hspace*{-2em}
\lim_{\rho\to0}\sum_{n=1}^{N}\big(\Sigma_{a}(E)\big|
\beta^{(+)}\mathcal{G}(E,\boldsymbol{r}')\big)_{\mathcal{S}_{n}}
\nonumber \\
& =\lim_{\rho\to0}\sum_{n=1}^{N}\big(\Sigma_{a}(E)\big|
\beta^{(+)}\mathcal{G}_{0}(E,\boldsymbol{r}')\big)_{\mathcal{S}_{n}}
+\sum_{b=1}^{2N}\left[\lim_{\rho\to0}
\sum_{n=1}^{N}\big(\Sigma_{a}(E)\big|
\beta^{(+)}\Sigma_{b}(E)\big)_{\mathcal{S}_{n}}\right]
C_{b}^{\dag}(E,\boldsymbol{r}').
\label{5.13}
\end{align}
The first term on the right-hand side of Eq.\ (\ref{5.13}) is zero,
whereas the second one simplifies after the orthonormality relation
(\ref{4.18}) is applied, yielding
\begin{equation}
\lim_{\rho\to0}\sum_{n=1}^{N}\big(\Sigma_{a}(E)\big|
\beta^{(+)}\mathcal{G}(E,\boldsymbol{r}')\big)_{\mathcal{S}_{n}}
=C_{a}^{\dag}(E,\boldsymbol{r}').
\label{5.14}
\end{equation}
Hence, one has
\begin{equation}
\lim_{\rho\to0}\big<\Sigma_{a}(E)\big|[\mathcal{H}-E\mathcal{I}]
\mathcal{G}(E,\boldsymbol{r}')\big>_{\mathbb{R}_{\rho}^{3}}
=-c\hbar\varepsilon\lambda_{a}(E)C_{a}^{\dag}(E,\boldsymbol{r}')
\label{5.15}
\end{equation}
and further, after Eq.\ (\ref{5.15}) is combined with Eq.\
(\ref{5.9}),
\begin{equation}
C_{a}^{\dag}(E,\boldsymbol{r}')
=-\frac{1}{c\hbar\varepsilon}\lambda_{a}^{-1}(E)
\Sigma_{a}^{\dag}(E,\boldsymbol{r}').
\label{5.16}
\end{equation}
Consequently, the sought form of the Sturmian representation
(\ref{5.4}) of $\mathcal{G}(E,\boldsymbol{r},\boldsymbol{r}')$ is
\begin{equation}
\mathcal{G}(E,\boldsymbol{r},\boldsymbol{r}')
=\mathcal{G}_{0}(E,\boldsymbol{r},\boldsymbol{r}')
-\frac{1}{c\hbar\varepsilon}
\sum_{a=1}^{2N}\lambda_{a}^{-1}(E)\Sigma_{a}(E,\boldsymbol{r})
\Sigma_{a}^{\dag}(E,\boldsymbol{r}').
\label{5.17}
\end{equation}
From Eqs.\ (\ref{5.17}) and (\ref{5.5}) the Green's function is seen
to be symmetric in the sense of
\begin{equation}
\mathcal{G}(E,\boldsymbol{r},\boldsymbol{r}')
=\mathcal{G}^{\dag}(E,\boldsymbol{r}',\boldsymbol{r}).
\label{5.18}
\end{equation}
%
%\newpage
%
\section{Illustrative applications}
\label{VI}
\setcounter{equation}{0}
\subsection{Particle bound in a field of a single zero-range
potential} 
\label{VI.1}
\subsubsection{Bound-state eigenenergies and associated
eigenfunctions}
\label{VI.1.1}
Consider a particle moving in a field of a single zero-range potential
located at the point $\boldsymbol{r}_{1}=\boldsymbol{0}$. The
$2\times2$ interaction matrix $K$ (for brevity, we omit the subscript
1) is
\begin{equation}
K=\varkappa I+\boldsymbol{\kappa}\cdot\boldsymbol{\sigma}.
\label{6.1}
\end{equation}
The matrix $\mathsf{L}(E_{a})$ is simply
\begin{equation}
\mathsf{L}(E_{a})=\frac{\hbar}{2mc\varepsilon_{a}}K-I
\label{6.2}
\end{equation}
and its determinant is
\begin{equation}
\det\mathsf{L}(E_{a})
=\left(\frac{\hbar\varkappa}{2mc\varepsilon_{a}}-1\right)^{2}
-\left(\frac{\hbar\kappa}{2mc\varepsilon_{a}}\right)^{2},
\label{6.3}
\end{equation}
where $\kappa=|\boldsymbol{\kappa}|$. Equating the right-hand side to
zero [cf.\ Eq.\ (\ref{2.19})] and solving the resulting equation for
$\varepsilon_{a}$ yields
\begin{equation}
\varepsilon_{\pm}=\frac{\hbar(\varkappa\pm\kappa)}{2mc},
\label{6.4}
\end{equation}
where we have put $a=\pm$ with reference to the two signs which appear
on the right-hand side. Since, by virtue of the definition
(\ref{2.4b}), $\varepsilon_{\pm}$ cannot be negative, we have the
following three possibilities:
\begin{equation}
\left\{
\begin{array}{ccl}
\varkappa<-\kappa & \Rightarrow & 
\textrm{there are no bound states} \\
-\kappa\leqslant\varkappa<\kappa & \Rightarrow & 
\textrm{there is one bound state of energy $E_{+}$} \\
\varkappa\geqslant\kappa & \Rightarrow & 
\textrm{there are two bound states of energies $E_{\pm}$},
\end{array}
\right.
\label{6.5}
\end{equation}
where
\begin{equation}
E_{\pm}=mc^{2}
\frac{\displaystyle
1-\left[\frac{\hbar(\varkappa\pm\kappa)}{2mc}\right]^{2}}
{\displaystyle
1+\left[\frac{\hbar(\varkappa\pm\kappa)}{2mc}\right]^{2}}.
\label{6.6}
\end{equation}
In accordance with the definition (\ref{2.4a}) or with the relation in
Eq.\ (\ref{2.6a}), wave numbers associated with the eigenenergies
(\ref{6.6}) are
\begin{equation}
k_{\pm}=\frac{\varkappa\pm\kappa}
{\displaystyle
1+\left[\frac{\hbar(\varkappa\pm\kappa)}{2mc}\right]^{2}}.
\label{6.7}
\end{equation}

It follows from what has been said above that in the limiting case
when $\boldsymbol{\kappa}=\boldsymbol{0}$ and $\varkappa\neq0$ (which
is the case of a `purely scalar' interaction), there are no bound
states if $\varkappa<0$, whereas if $\varkappa>0$, then there are two
degenerate bound states of energy
\begin{equation}
E=mc^{2}
\frac{\displaystyle1-\left(\frac{\hbar\varkappa}{2mc}\right)^{2}}
{\displaystyle1+\left(\frac{\hbar\varkappa}{2mc}\right)^{2}},
\label{6.8}
\end{equation}
the redundant subscript at $E$ being omitted. In the second limiting
case, i.e., for $\varkappa=0$ and
$\boldsymbol{\kappa}\neq\boldsymbol{0}$ (which is the case of a
`purely vector' interaction), there will always be only one bound
state of energy
\begin{equation}
E_{+}=mc^{2}
\frac{\displaystyle1-\left(\frac{\hbar\kappa}{2mc}\right)^{2}}
{\displaystyle1+\left(\frac{\hbar\kappa}{2mc}\right)^{2}}.
\label{6.9}
\end{equation}

In the nonrelativistic approximation, Eqs.\ (\ref{6.6}) and
(\ref{6.7}) go over into
\begin{equation}
E_{\pm}\simeq mc^{2}-\frac{\hbar^{2}(\varkappa\pm\kappa)^{2}}{2m}
\label{6.10}
\end{equation}
and
\begin{equation}
k_{\pm}=\varkappa\pm\kappa,
\label{6.11}
\end{equation}
respectively. Approximations analogous to that in Eq.\ (\ref{6.10})
obviously hold for Eqs.\ (\ref{6.8}) and (\ref{6.9}).

Next, we turn to the eigenfunctions. Adapting Eq.\ (\ref{2.2b}) to the
present case, we see that the eigenfunctions which belong to the
eigenenergies $E_{\pm}$ are
\begin{equation}
\Psi_{\pm}(\boldsymbol{r})
=\left(
\begin{array}{c}
f(k_{\pm}r)\chi_{\pm} \\*[1ex]
\mathrm{i}\varepsilon_{\pm}
g(k_{\pm} r)\boldsymbol{n}_{r}\cdot\boldsymbol{\sigma}\chi_{\pm}
\end{array}
\right),
\label{6.12}
\end{equation}
where $\boldsymbol{n}_{r}=\boldsymbol{r}/r$ is the unit vector in the
direction of the position vector $\boldsymbol{r}$ and where the spinor
coefficients $\chi_{\pm}$ solve [cf.\ Eq.\ (\ref{2.14})]
\begin{equation}
\left(\frac{\hbar}{2mc\varepsilon_{\pm}}K-I\right)\chi_{\pm}=0.
\label{6.13}
\end{equation}
Invoking Eq.\ (\ref{6.1}), it is easy to see that the spinors
$\chi_{\pm}$ may be written as
\begin{equation}
\chi_{\pm}=c_{\pm}\xi_{\pm},
\label{6.14}
\end{equation}
where $c_{\pm}$ are normalization coefficients to be determined later,
whereas $\xi_{\pm}$ are normalized (in the sense of
$\xi_{\pm}^{\dag}\xi_{\pm}=1$) eigenvectors of the matrix
$\boldsymbol{\kappa}\cdot\boldsymbol{\sigma}$ and obey
\begin{equation}
\boldsymbol{\kappa}\cdot\boldsymbol{\sigma}\xi_{\pm}
=\pm\kappa\xi_{\pm}.
\label{6.15}
\end{equation}
The explicit forms of the spinors $\xi_{\pm}$ are
\begin{equation}
\xi_{+}
=\left(
\begin{array}{c}
\cos(\theta_{\kappa}/2) \\
\sin(\theta_{\kappa}/2)\mathrm{e}^{\mathrm{i}\phi_{\kappa}}
\end{array}
\right),
\qquad
\xi_{-}
=\left(
\begin{array}{c}
\sin(\theta_{\kappa}/2) \\
-\cos(\theta_{\kappa}/2)\mathrm{e}^{\mathrm{i}\phi_{\kappa}}
\end{array}
\right),
\label{6.16}
\end{equation}
where $0\leqslant\theta_{\kappa}\leqslant\pi$ and
$0\leqslant\phi_{\kappa}<2\pi$ are the spherical angles of the vector
$\boldsymbol{\kappa}$.

In Sec.\ \ref{III}, we have mentioned that the single-center system
considered here is the one for which the eigenfunctions may be
effectively normalized using any of the two available representations
of the self-pseudo-product. To show that this is indeed the case,
consider at first the pseudo-product
$\langle\!\langle\Psi_{\pm}\big|\Psi_{\pm}\rangle\!\rangle$ in the
form (\ref{3.17}), i.e.,
\begin{equation}
\langle\!\langle\Psi_{\pm}\big|\Psi_{\pm}\rangle\!\rangle
=\lim_{\rho\to0}
\left\{\big<\Psi_{\pm}\big|\Psi_{\pm}\big>_{\mathbb{R}_{\rho}^{3}}
-\frac{\varepsilon_{\pm}^{2}k_{\pm}^{-2}}{\rho}
\big(\Psi_{\pm}\big|
\beta^{(+)}\Psi_{\pm}\big)_{\mathcal{S}}\right\}
\label{6.17}
\end{equation}
(the redundant subscript at $\mathcal{S}$ has been omitted
intentionally). With no difficulty one finds that
\begin{subequations}
\begin{equation}
\big<\Psi_{\pm}\big|\Psi_{\pm}\big>_{\mathbb{R}_{\rho}^{3}}
=\frac{2\pi}{k_{\pm}^{3}}\left[1+\varepsilon_{\pm}^{2}
\left(1+\frac{2}{k_{\pm}\rho}\right)\right]\mathrm{e}^{-2k_{\pm}\rho}
\chi_{\pm}^{\dag}\chi_{\pm}
\label{6.18a}
\end{equation}
and
\begin{equation}
\big(\Psi_{\pm}\big|\beta^{(+)}\Psi_{\pm}\big)_{\mathcal{S}}
=\frac{4\pi}{k_{\pm}^{2}}\mathrm{e}^{-2k_{\pm}\rho}
\chi_{\pm}^{\dag}\chi_{\pm},
\label{6.18b}
\end{equation}
\label{6.18}%
\end{subequations}
and consequently one has
\begin{equation}
\lim_{\rho\to0}
\left\{\big<\Psi_{\pm}\big|\Psi_{\pm}\big>_{\mathbb{R}_{\rho}^{3}}
-\frac{\varepsilon_{\pm}^{2}k_{\pm}^{-2}}{\rho}
\big(\Psi_{\pm}\big|\beta^{(+)}\Psi_{\pm}\big)_{\mathcal{S}}\right\}
=\frac{2\pi}{k_{\pm}^{3}}\big(1+\varepsilon_{\pm}^{2}\big)
\chi_{\pm}^{\dag}\chi_{\pm}.
\label{6.19}
\end{equation}
Alternatively, we may take the pseudo-product
$\langle\!\langle\Psi_{\pm}\big|\Psi_{\pm}\rangle\!\rangle$ in the
form
\begin{equation}
\langle\!\langle\Psi_{\pm}\big|\Psi_{\pm}\rangle\!\rangle
=\big(1-\varepsilon_{\pm}^{2}\big)
\big<\Psi_{\pm}\big|\beta^{(+)}\Psi_{\pm}\big>_{\mathbb{R}^{3}}
+\frac{\hbar}{2mc}\varepsilon_{\pm}k_{\pm}^{-1}
\lim_{\rho\to0}\big(\Psi_{\pm}\big|
\mathcal{K}^{(+)}\Psi_{\pm}\big)_{\mathcal{S}},
\label{6.20}
\end{equation}
which follows from Eq.\ (\ref{3.31}). For the two integrals involved
one easily obtains
\begin{subequations}
\begin{equation}
\big<\Psi_{\pm}\big|\beta^{(+)}\Psi_{\pm}\big>_{\mathbb{R}^{3}}
=\frac{2\pi}{k_{\pm}^{3}}\chi_{\pm}^{\dag}\chi_{\pm}
\label{6.21a}
\end{equation}
and
\begin{equation}
\big(\Psi_{\pm}\big|\mathcal{K}^{(+)}\Psi_{\pm}\big)_{\mathcal{S}}
=\frac{4\pi(\varkappa\pm\kappa)}{k_{\pm}^{2}}
\mathrm{e}^{-2k_{\pm}\rho}\chi_{\pm}^{\dag}\chi_{\pm},
\label{6.21b}
\end{equation}
\label{6.21}%
\end{subequations}
respectively. This brings the right-hand side of Eq.\ (\ref{6.20}) to
the form
\begin{equation}
\big(1-\varepsilon_{\pm}^{2}\big)
\big<\Psi_{\pm}\big|\beta^{(+)}\Psi_{\pm}\big>_{\mathbb{R}^{3}}
+\frac{\hbar}{2mc}\varepsilon_{\pm}k_{\pm}^{-1}\lim_{\rho\to0}
\big(\Psi_{\pm}\big|\mathcal{K}^{(+)}\Psi_{\pm}\big)_{\mathcal{S}}
=\frac{2\pi}{k_{\pm}^{3}}\big(1+\varepsilon_{\pm}^{2}\big)
\chi_{\pm}^{\dag}\chi_{\pm}.
\label{6.22}
\end{equation}
We thus see that no matter which of the two available representations
of $\langle\!\langle\Psi_{\pm}\big|\Psi_{\pm}\rangle\!\rangle$ is
used, one gets
\begin{equation}
\langle\!\langle\Psi_{\pm}\big|\Psi_{\pm}\rangle\!\rangle
=\frac{2\pi}{k_{\pm}^{3}}\big(1+\varepsilon_{\pm}^{2}\big)
\chi_{\pm}^{\dag}\chi_{\pm}.
\label{6.23}
\end{equation}
The right-hand side of Eq.\ (\ref{6.23}) is positive and this implies
that the signatures of the eigenfunctions $\Psi_{\pm}(\boldsymbol{r})$
are
\begin{equation}
\Delta_{\pm}=+1.
\label{6.24}
\end{equation}
Hence, the explicit form of the normalization condition (\ref{3.21})
is
\begin{equation}
\frac{2\pi}{k_{\pm}^{3}}\big(1+\varepsilon_{\pm}^{2}\big)
\chi_{\pm}^{\dag}\chi_{\pm}=1.
\label{6.25}
\end{equation}
On combining Eq.\ (\ref{6.25}) with the relation
\begin{equation}
\chi_{\pm}^{\dag}\chi_{\pm}=|c_{\pm}|^{2},
\label{6.26}
\end{equation}
which follows from Eq.\ (\ref{6.14}) and from the unitary
normalization of $\xi_{\pm}$, one arrives at the result
\begin{equation}
c_{\pm}=\sqrt{\frac{k_{\pm}^{3}}
{2\pi\big(1+\varepsilon_{\pm}^{2}\big)}}.
\label{6.27}
\end{equation}
For convenience, an indeterminable phase factor has been chosen to
make $c_{\pm}$ real and positive. Hence, the normalized eigenfunctions
are
\begin{equation}
\Psi_{\pm}(\boldsymbol{r})
=\sqrt{\frac{k_{\pm}^{3}}{2\pi\big(1+\varepsilon_{\pm}^{2}\big)}}
\left(
\begin{array}{c}
f(k_{\pm}r)\xi_{\pm} \\*[1ex]
\mathrm{i}\varepsilon_{\pm}
g(k_{\pm}r)\boldsymbol{n}_{r}\cdot\boldsymbol{\sigma}\xi_{\pm}
\end{array}
\right),
\label{6.28}
\end{equation}
with the caveat that if the only energy eigenvalue is $E_{+}$, then
only $\Psi_{+}(\boldsymbol{r})$ is a physically meaningful
eigenfunction.
\subsubsection{The Sturmian functions}
\label{VI.1.2}
To construct the Sturmian functions (\ref{4.2}) for the system under
consideration, we have to solve the eigenvalue problem
\begin{equation}
\mathsf{L}(E)\eta_{a}(E)=\lambda_{a}(E)\eta_{a}(E)
\label{6.29}
\end{equation}
with the matrix $\mathsf{L}(E)$ given by
\begin{equation}
\mathsf{L}(E)=\frac{\hbar}{2mc\varepsilon}K-I.
\label{6.30}
\end{equation}
The Sturmian eigenvalues, i.e., the roots of the characteristic
equation
\begin{equation}
\det[\mathsf{L}(E)-\lambda_{a}(E)I]=0,
\label{6.31}
\end{equation}
are
\begin{equation}
\lambda_{\pm}(E)=\frac{\hbar(\varkappa\pm\kappa)}{2mc\varepsilon}-1
=\frac{\varepsilon_{\pm}}{\varepsilon}-1,
\label{6.32}
\end{equation}
with $\varepsilon_{\pm}$ defined in Eq.\ (\ref{6.4}). Associated
spinor coefficients $\eta_{\pm}(E)$, normalized in accordance with
\begin{equation}
\eta_{\pm}^{\dag}(E)\eta_{\pm}(E)=\frac{k^{2}}{4\pi}
\label{6.33}
\end{equation}
and suitably phased, are then found to be
\begin{equation}
\eta_{\pm}(E)=\frac{k}{\sqrt{4\pi}}\xi_{\pm}.
\label{6.34}
\end{equation}
This yields the sought Sturmian functions in the form
\begin{equation}
\Sigma_{\pm}(E,\boldsymbol{r})
=\frac{k}{\sqrt{4\pi}}
\left(
\begin{array}{c}
f(kr)\xi_{\pm} \\*[1ex]
\mathrm{i}\varepsilon g(kr)
\boldsymbol{n}_{r}\cdot\boldsymbol{\sigma}\xi_{\pm}
\end{array}
\right).
\label{6.35}
\end{equation}
Since it holds that
\begin{equation}
\left[\frac{\partial\lambda_{\pm}(E)}{\partial E}\right]_{E=E_{\pm}}
=\frac{m}{\hbar^{2}k_{\pm}^{2}},
\label{6.36}
\end{equation}
with $E_{\pm}$ and $k_{\pm}$ defined in Eqs.\ (\ref{6.6}) and
(\ref{6.7}), respectively, upon exploiting Eq.\ (\ref{4.27}) we arrive
at the relationship
\begin{equation}
\Psi_{\pm}(\boldsymbol{r})
=\sqrt{\frac{2k_{\pm}}{1+\varepsilon_{\pm}^{2}}}\,
\Sigma_{\pm}(E_{\pm},\boldsymbol{r}),
\label{6.37}
\end{equation}
which, by virtue of Eq.\ (\ref{6.35}), is seen to be in agreement with
the result in Eq.\ (\ref{6.28}).
%
%\newpage
%
\subsection{Particle bound in a field of two identical zero-range
potentials} 
\label{VI.2}
\begin{figure}[t!]
\begin{center}
%\hspace*{-1em}
\includegraphics[width=1\textwidth]{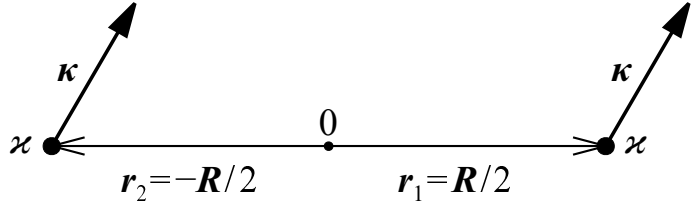}
\end{center}
\vspace*{-10ex}
\caption{Two identical zero-range potentials located at the points
$\boldsymbol{r}_{1}=\boldsymbol{R}/2$ and
$\boldsymbol{r}_{2}=-\boldsymbol{R}/2$, respectively. Each potential
is characterized by the real scalar $\varkappa$ and the real vector
$\boldsymbol{\kappa}$, which together define the $2\times2$
interaction matrix $K$ in accordance with Eq.\ (\ref{6.1}).}
\label{FIG2}
\end{figure}

As the second example, let us consider a particle bound in the field
of two identical zero-range potentials which are located at the points
\begin{equation}
\boldsymbol{r}_{1}=\frac{1}{2}\boldsymbol{R},
\qquad
\boldsymbol{r}_{2}=-\frac{1}{2}\boldsymbol{R},
\label{6.38}
\end{equation}
respectively (cf.\ Fig.\ \ref{FIG2}), and are characterized by the
$2\times2$ interaction matrices
\begin{equation}
K_{1}=K_{2}=K,
\label{6.39}
\end{equation}
with $K$ defined as in Eq.\ (\ref{6.1}). This time we shall consider
the relevant Sturmian problem first and then proceed to the analysis
of the energy eigenproblem.
\subsubsection{The Sturmian functions}
\label{VI.2.1}
The matrix $\mathsf{L}(E)$ for the system under study is
\begin{equation}
\mathsf{L}(E)
=\left(
\begin{array}{cc}
{\displaystyle\frac{\hbar}{2mc\varepsilon}
(\varkappa I+\boldsymbol{\kappa}\cdot\boldsymbol{\sigma})-I} & 
\displaystyle{\frac{\mathrm{e}^{-kR}}{kR}I} \\*[2ex]
\displaystyle{\frac{\mathrm{e}^{-kR}}{kR}I} & 
\displaystyle{\frac{\hbar}{2mc\varepsilon}
(\varkappa I+\boldsymbol{\kappa}\cdot\boldsymbol{\sigma})-I}
\end{array}
\right).
\label{6.40}
\end{equation}
It is not difficult to show that its four eigenvalues are
\begin{subequations}
\begin{align}
\lambda_{\pm g}(E)
& =\frac{\hbar(\varkappa\pm\kappa)}{2mc\varepsilon}-1
+\frac{\mathrm{e}^{-kR}}{kR},
\label{6.41a}
\\
\lambda_{\pm u}(E)
& =\frac{\hbar(\varkappa\pm\kappa)}{2mc\varepsilon}-1
-\frac{\mathrm{e}^{-kR}}{kR},
\label{6.41b}
\end{align}
\label{6.41}%
\end{subequations}
and that the associated eigenvectors normalized in accordance with
Eq.\ (\ref{4.19}) are
\begin{subequations}
\begin{align}
\mathsf{y}_{\pm g}(E)
& =\frac{k}{\sqrt{8\pi}}
\left(
\begin{array}{cc}
\xi_{\pm}^{\mathrm{T}} & \xi_{\pm}^{\mathrm{T}}
\end{array}
\right)^{\mathrm{T}},
\label{6.42a}
\\
\mathsf{y}_{\pm u}(E)
& =\frac{k}{\sqrt{8\pi}}
\left(
\begin{array}{cc}
\xi_{\pm}^{\mathrm{T}} & -\xi_{\pm}^{\mathrm{T}}
\end{array}
\right)^{\mathrm{T}},
\label{6.42b}
\end{align}
\label{6.42}%
\end{subequations}
where $\xi_{\pm}$ are the eigenvectors of the matrix
$\boldsymbol{\kappa}\cdot\boldsymbol{\sigma}$ displayed in Eq.\
(\ref{6.16}). Hence, the Sturmian functions for the current problem
are given by
\begin{subequations}
\begin{align}
\Sigma_{\pm g}(E,\boldsymbol{r})
& =\frac{k}{\sqrt{8\pi}}
\left(
\begin{array}{c}
\big[f(k|\boldsymbol{r}-\boldsymbol{R}/2|)
+f(k|\boldsymbol{r}+\boldsymbol{R}/2|)\big]\xi_{\pm}
\\*[1ex]
\mathrm{i}\varepsilon\big[g(k|\boldsymbol{r}-\boldsymbol{R}/2|)
\boldsymbol{\mu}(\boldsymbol{r},\boldsymbol{R}/2)
+g(k|\boldsymbol{r}+\boldsymbol{R}/2|)
\boldsymbol{\mu}(\boldsymbol{r},-\boldsymbol{R}/2)\big]
\cdot\boldsymbol{\sigma}\xi_{\pm}
\end{array}
\right),
\label{6.43a}
\end{align}
\begin{align}
\Sigma_{\pm u}(E,\boldsymbol{r})
& =\frac{k}{\sqrt{8\pi}}
\left(
\begin{array}{c}
\big[f(k|\boldsymbol{r}-\boldsymbol{R}/2|)
-f(k|\boldsymbol{r}+\boldsymbol{R}/2|)\big]\xi_{\pm}
\\*[1ex]
\mathrm{i}\varepsilon\big[g(k|\boldsymbol{r}-\boldsymbol{R}/2|)
\boldsymbol{\mu}(\boldsymbol{r},\boldsymbol{R}/2)
-g(k|\boldsymbol{r}+\boldsymbol{R}/2|)
\boldsymbol{\mu}(\boldsymbol{r},-\boldsymbol{R}/2)\big]
\cdot\boldsymbol{\sigma}\xi_{\pm}
\end{array}
\right),
\label{6.43b}
\end{align}
\label{6.43}%
\end{subequations}
[for the definition of the unit vectors
$\boldsymbol{\mu}(\boldsymbol{r},\pm\boldsymbol{R}/2)$ see Eq.\
(\ref{5.6})]. It is evident that the Sturmian functions with the
subscript $g$ (respectively, $u$) are eigenfunctions of the Dirac
parity operator $\Pi$ [defined through its action on an arbitrary
bispinor function $\Phi(\boldsymbol{r})$ in the following way:
$\Pi\Phi(\boldsymbol{r})\equiv\beta\Phi(-\boldsymbol{r})$, where
$\beta$ is the Dirac beta matrix] associated with the eigenvalue $+1$
(respectively, $-1$).
\subsubsection{Bound-state eigenenergies}
\label{VI.2.2}
\begin{figure}[t!]
\begin{center}
\includegraphics[width=1\textwidth]{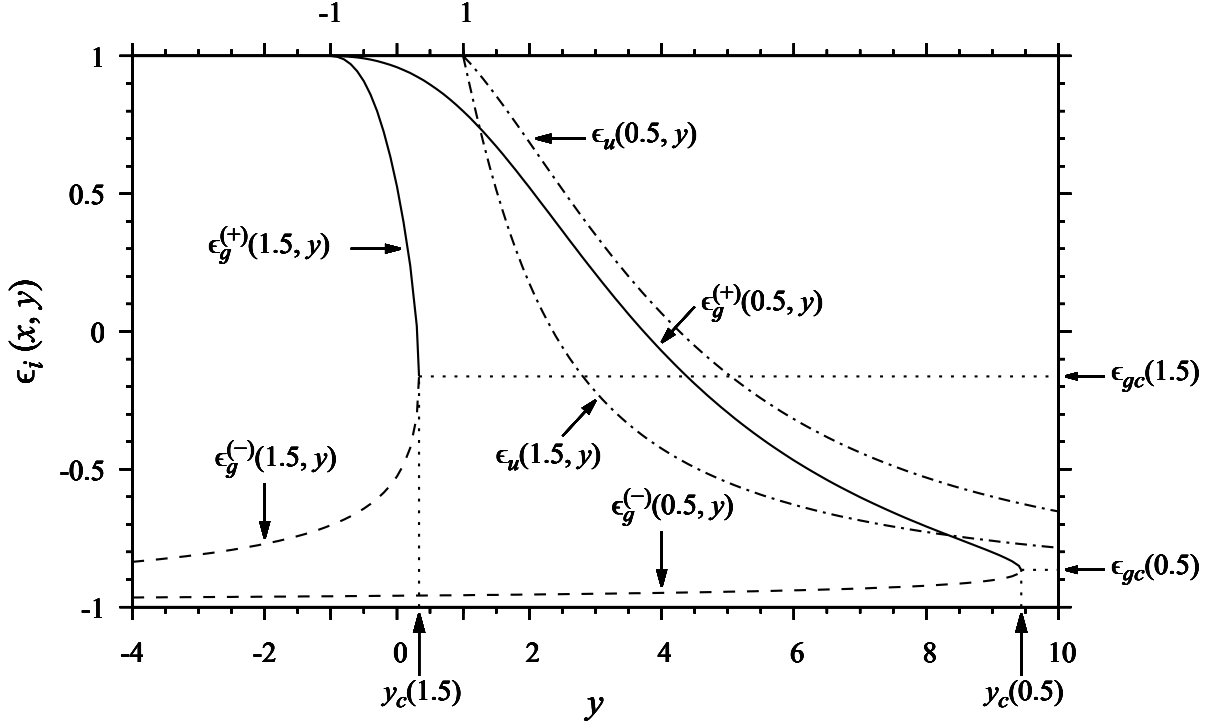}
\end{center}
\vspace*{-39ex}
\caption{Plots of the functions $\epsilon_{g}(x,y)$ and
$\epsilon_{u}(x,y)$ [being real solutions to Eqs.\ (\ref{6.46a}) and
(\ref{6.46b}), respectively] versus $y$ for $x=0.5$ and for $x=1.5$.
The function $\epsilon_{u}(x,y)$ is single-valued. The function
$\epsilon_{g}(x,y)$ has two branches, denoted as
$\epsilon_{g}^{(+)}(x,y)$ and $\epsilon_{g}^{(-)}(x,y)$, which match
smoothly, with an infinite slope, at the point
$\{y_{c}(x),\epsilon_{gc}(x)\}$. One has
$\{y_{c}(0.5)=9.437...,\epsilon_{gc}(0.5)=-0.865...\}$ and
$\{y_{c}(1.5)=0.334...,\epsilon_{gc}(1.5)=-0.163...\}$ (for more
precise values of the $y_{c}$'s and $\epsilon_{gc}$'s, see Table
\ref{TAB2}).}
\label{FIG3}
\end{figure}

Algebraic equations leading to particle's energy eigenvalues are
obtained by equating each of the Sturmian eigenvalues (\ref{6.41}) to
zero:
\begin{subequations}
\begin{align}
\lambda_{\pm g}(E_{\pm g})
&\equiv\frac{\hbar(\varkappa\pm\kappa)}{2mc}
\sqrt{\frac{mc^{2}+E_{\pm g}}{mc^{2}-E_{\pm g}}}-1+\frac{c\hbar}{R}
\frac{\mathrm{e}^{-\sqrt{(mc^{2})^{2}-E_{\pm g}^{2}}(R/c\hbar)}}
{\sqrt{(mc^{2})^{2}-E_{\pm g}^{2}}}=0,
\label{6.44a}
\\
\lambda_{\pm u}(E_{\pm u})
& \equiv\frac{\hbar(\varkappa\pm\kappa)}{2mc}
\sqrt{\frac{mc^{2}+E_{\pm u}}{mc^{2}-E_{\pm u}}}-1-\frac{c\hbar}{R}
\frac{\mathrm{e}^{-\sqrt{(mc^{2})^{2}-E_{\pm u}^{2}}(R/c\hbar)}}
{\sqrt{(mc^{2})^{2}-E_{\pm u}^{2}}}=0.
\label{6.44b}
\end{align}
\label{6.44}%
\end{subequations}
It is not difficult to see that roots to Eqs.\ (\ref{6.44}) may be
expressed in the following manner:
\begin{subequations}
\begin{align}
E_{\pm g} & =mc^{2}
\epsilon_{g}\left(\frac{\hbar}{mcR},(\varkappa\pm\kappa)R\right),
\label{6.45a}
\\
E_{\pm u} & =mc^{2}
\epsilon_{u}\left(\frac{\hbar}{mcR},(\varkappa\pm\kappa)R\right)
\label{6.45b}
\end{align}
\label{6.45}%
\end{subequations}
in terms of two universal functions
$\epsilon_{g}(x,y):\mathbb{R}_{+}\times\mathbb{R}\to(-1,+1]$ and
$\epsilon_{u}(x,y):\mathbb{R}_{+}\times\mathbb{R}\to(-1,+1]$, which
are solutions to the transcendental algebraic equations
\begin{subequations}
\begin{equation}
\frac{1}{2}xy\sqrt{\frac{1+\epsilon_{g}(x,y)}{1-\epsilon_{g}(x,y)}}-1
+\frac{x\exp\left(-\frac{1}{x}\sqrt{1-\epsilon_{g}^{2}(x,y)}\right)}
{\sqrt{1-\epsilon_{g}^{2}(x,y)}}=0
\label{6.46a}
\end{equation}
and
\begin{equation}
\frac{1}{2}xy\sqrt{\frac{1+\epsilon_{u}(x,y)}{1-\epsilon_{u}(x,y)}}-1
-\frac{x\exp\left(-\frac{1}{x}\sqrt{1-\epsilon_{u}^{2}(x,y)}\right)}
{\sqrt{1-\epsilon_{u}^{2}(x,y)}}=0,
\label{6.46b}
\end{equation}
\label{6.46}%
\end{subequations}
respectively. Equations (\ref{6.46}) define $\epsilon_{g}(x,y)$ and
$\epsilon_{u}(x,y)$ in an implicit manner. Explicit algebraic
representations of the two functions remain unknown and to make graphs
of $\epsilon_{g}(x,y)$ and $\epsilon_{u}(x,y)$, one has to solve Eqs.\
(\ref{6.46}) numerically. We have done this with Mathematica 12.3.
Representative plots, obtained for two fixed values of $x$ and for
varying $y$, are depicted in Fig.\ \ref{FIG3}. It is seen that
behaviors of the two functions are completely different. The function
$\epsilon_{u}(x,y)$ exists for $y\geqslant1$. It is single-valued and
decreases monotonically from $\epsilon_{u}(x,1)=1$ to
$\lim_{y\to\infty}\epsilon_{u}(x,y)=-1$. To the contrary,
$\epsilon_{g}(x,y)$ is a two-branched function. The branch
$\epsilon_{g}^{(+)}(x,y)$, which exists for $-1\leqslant y\leqslant
y_{c}(x)$, decreases monotonically from $\epsilon_{g}^{(+)}(x,-1)=1$
to $\epsilon_{g}^{(+)}\big(x,y_{c}(x)\big)=\epsilon_{gc}(x)$, with
$[\partial\epsilon_{g}^{(+)}(x,y)/\partial y]_{y=y_{c}(x)}=-\infty$.
The branch $\epsilon_{g}^{(-)}(x,y)$, which exists for
$-\infty<y\leqslant y_{c}(x)$, increases monotonically from
$\lim_{y\to-\infty}\epsilon_{g}^{(-)}(x,y)=-1$ to
$\epsilon_{g}^{(-)}\big(x,y_{c}(x)\big)=\epsilon_{gc}(x)$, with
$[\partial\epsilon_{g}^{(-)}(x,y)/\partial y]_{y=y_{c}(x)}=\infty$.
The two branches match smoothly at the point
$\{y_{c}(x),\epsilon_{gc}(x)\}$. Hence, the function
$\epsilon_{g}(x,y)$ is single-valued in the interval $-\infty<y<-1$
and at the point $y=y_{c}(x)$, being double-valued in the interval
$-1\leqslant y<y_{c}(x)$. The coordinates of the matching point
$\{y_{c}(x),\epsilon_{gc}(x)\}$ may be found by solving the algebraic
system
\begin{equation}
\left\{
\begin{array}{l}
\displaystyle\frac{1}{2}xy_{c}(x)
\sqrt{\frac{1+\epsilon_{gc}(x)}{1-\epsilon_{gc}(x)}}-1
+\frac{x\exp\left(-\frac{1}{x}\sqrt{1-\epsilon_{gc}^{2}(x)}\right)}
{\sqrt{1-\epsilon_{gc}^{2}(x)}}=0, \\*[3ex]
\displaystyle\frac{1}{2}xy_{c}(x)
\sqrt{\frac{1+\epsilon_{gc}(x)}{1-\epsilon_{gc}(x)}}
+\epsilon_{gc}(x)
\left(1+\frac{x}{\sqrt{1-\epsilon_{gc}^{2}(x)}}\right)
\exp\left(-\frac{1}{x}\sqrt{1-\epsilon_{gc}^{2}(x)}\right)=0.
\end{array}
\right.
\label{6.47}
\end{equation}
The first equation in this system follows from the fact that the pair
$\{y_{c}(x),\epsilon_{gc}(x)\}$ has to obey Eq.\ (\ref{6.46a}). The
second one is the consequence of the fact that at the matching point
the slope of $\epsilon_{g}(x,y)$ versus $y$ is infinite [cf.\ the text
preceding Eq.\ (\ref{6.47})]. Its explicit form results after Eq.\
(\ref{6.46a}) is differentiated with respect to $y$, the resulting
identity is divided by $\partial\epsilon_{g}(x,y)/\partial y$ and then
the constraint $\partial\epsilon_{g}(x,y)/\partial y=\pm\infty$ is
imposed for $y=y_{c}(x)$ and
$\epsilon_{g}\big(x,y_{c}(x)\big)=\epsilon_{gc}(x)$.

Although exact analytical representations of
$\epsilon_{g}^{(\pm)}(x,y)$ and $\epsilon_{u}(x,y)$ are not available,
employing iteration methods we have been able to derive the following
truncated-series approximations to these functions:
\begin{align}
\epsilon_{g}^{(+)}(x,y) & \stackrel{x\to0+0}{\sim}
1-\frac{x^{2}}{2}\left[y+W_{0}\big(\mathrm{e}^{-y}\big)\right]^{2}
\nonumber \\*
& \quad\qquad +\frac{x^{4}}{8}
\frac{\big[y+W_{0}\big(\mathrm{e}^{-y}\big)\big]^{3}}
{1+W_{0}\big(\mathrm{e}^{-y}\big)}
\Big\{y-(y+1)W_{0}\big(\mathrm{e}^{-y}\big)
-\big[W_{0}\big(\mathrm{e}^{-y}\big)\big]^{2}\Big\}+O\big(x^6\big),
\label{6.48}
\end{align}
\begin{equation}
\epsilon_{g}^{(+)}(x,y)\stackrel{y\to-1+0}{\sim}
1-\frac{1}{8}x^{2}(y+1)^2-\frac{1}{64}x^{2}(x^2+2)(y+1)^3
+O\big((y+1)^4\big),
\label{6.49}
\end{equation}
\begin{align}
& \epsilon_{g}^{(-)}(x,y)\stackrel{y\to-\infty}{\sim}-1+\frac{2}{|y|}
-\frac{8}{x|y|^{3/2}}+\frac{20}{x^2|y|^{2}}
+\frac{4(3x^{2}-32)}{3x^{3}|y|^{5/2}}
-\frac{4(27x^{2}-71)}{3x^{4}|y|^{3}}+O\big(|y|^{-7/2}\big),
\label{6.50}
\end{align}
\begin{align}
\epsilon_{u}(x,y) & \stackrel{x\to0+0}{\sim}1-\frac{x^{2}}{2}
\left[y+W_{0}\big(\!-\!\mathrm{e}^{-y}\big)\right]^{2}
\nonumber \\*
& \quad\qquad +\frac{x^{4}}{8}
\frac{\big[y+W_{0}\big(\!-\!\mathrm{e}^{-y}\big)\big]^{3}}
{1+W_{0}\big(\!-\!\mathrm{e}^{-y}\big)}
\Big\{y-(y+1)W_{0}\big(\!-\!\mathrm{e}^{-y}\big)
-\big[W_{0}\big(\!-\!\mathrm{e}^{-y}\big)\big]^{2}\Big\}
+O\big(x^{6}\big),
\label{6.51}
\end{align}
\begin{align}
\epsilon_{u}(x,y) & \stackrel{y\to1+0}{\sim}
1-2\frac{x^2}{x^{2}+2}(y-1)
-\frac{8}{3}\frac{x^{2}}{(x^{2}+2)^{5/2}}(y-1)^{3/2}
+\frac{2}{3}\frac{x^{2}(3x^{6}+6x^{4}+2x^{2}-4)}{(x^{2}+2)^4}(y-1)^{2}
\nonumber \\*
& \quad\qquad +O\big((y-1)^{5/2}\big),
\label{6.52}
\end{align}
\begin{equation}
\epsilon_{u}(x,y)\stackrel{y\to\infty}{\sim}
-1+\frac{2}{y}+\frac{4}{x^{2}y^{2}}-\frac{8}{3x^{3}y^{5/2}}
-\frac{4(3x^{2}-7)}{3x^{4}y^{3}}+O\big(y^{-7/2}\big).
\label{6.53}
\end{equation}
In Eqs.\ (\ref{6.48}) and (\ref{6.51}), $W_{0}(z)$ denotes the
principal branch of the Lambert function (the product logarithm)
\cite{Corl96,Dubi06}.
\begin{table}[t!]
\caption{The table marks existence ($\surd$) or non-existence (---) of
the solutions $\epsilon_{g}^{(-)}(x,y)$ and $\epsilon_{g}^{(+)}(x,y)$
to Eq.\ (\ref{6.46a}) and the solution $\epsilon_{u}(x,y)$ to Eq.\
(\ref{6.46b}) for various combinations of subdomains that $x$ and $y$
may belong to. For $y=y_{c}(x)$, the equality sign placed between the
second and the third columns reminds that
$\epsilon_{g}^{(-)}\big(x,y_{c}(x)\big)
=\epsilon_{g}^{(+)}\big(x,y_{c}(x)\big)=\epsilon_{gc}(x)$. For the
critical value of $x=x_{c}=1.198\,076...$, one has $y=y_{c}(x_{c})=1$
and $\epsilon_{g}^{(-)}(x_{c},1)=\epsilon_{g}^{(+)}(x_{c},1)
=\epsilon_{gc}(x_{c})=-0.379\,162...$.}
\label{TAB1}
\begin{center}
\begin{tabular}{r@{}c@{}lcccccc}
\hline\hline \\*[0ex]
\multicolumn{3}{c}{Range of $y$} && $\epsilon_{g}^{(-)}(x,y)$ && 
$\epsilon_{g}^{(+)}(x,y)$ && $\epsilon_{u}(x,y)$ \\*[2ex] 
\hline \\*[0ex]
\multicolumn{9}{c}{$0<x<x_{c}$} \\*[2ex]
$-\infty<$ & $\;y\;$ & $<-1$ && 
$\surd$ && --- && --- \\*[1ex]
$-1\leqslant$ & $y$ & $<1$ && 
$\surd$ && $\surd$ && --- \\*[1ex]
$1\leqslant$ & $y$ & $<y_{c}(x)$ && 
$\surd$ && $\surd$ && $\surd$ \\*[1ex]
& $y$ & $=y_{c}(x)$ &&
$\surd$ &=& $\surd$ && $\surd$ \\*[1ex]
$y_{c}(x)<$ & $y$ & $<\infty$ &&
--- && --- && $\surd$ \\*[2ex]
\multicolumn{9}{c}{$x=x_{c}=1.198\,076...$} \\*[2ex]
$-\infty<$ & $\;y\;$ & $<-1$ && 
$\surd$ && --- && --- \\*[1ex]
$-1\leqslant$ & $y$ & $<y_{c}(x_{c})=1$ &&
$\surd$ && $\surd$ && --- \\*[1ex]
& $y$ & $=y_{c}(x_{c})=1$ &&
$\surd$ &=& $\surd$ && $\surd$ \\*[1ex]
$1=y_{c}(x_{c})<$ & $y$ & $<\infty$ &&
--- && --- && $\surd$ \\*[2ex]
\multicolumn{9}{c}{$x_{c}<x<\infty$} \\*[2ex]
$-\infty<$ & $\;y\;$ & $<-1$ && 
$\surd$ && --- && --- \\*[1ex]
$-1\leqslant$ & $y$ & $<y_{c}(x)$ &&
$\surd$ && $\surd$ && --- \\*[1ex]
& $y$ & $=y_{c}(x)$ &&
$\surd$ &=& $\surd$ && --- \\*[1ex]
$y_{c}(x)<$ & $y$ & $<1$ && 
--- && --- && --- \\*[1ex]
$1\leqslant$ & $y$ & $<\infty$ &&
--- && --- && $\surd$ \\*[2ex]
\hline\hline
\end{tabular}
\end{center}
\end{table}
\begin{table}[t!]
\caption{Numerical values of the solutions $\epsilon_{g}^{(-)}(x,y)$ and
$\epsilon_{g}^{(+)}(x,y)$ to Eq.\ (\ref{6.46a}) and of the solution
$\epsilon_{u}(x,y)$ to Eq.\ (\ref{6.46b}) for selected values of $x$
and $y$. For $y=y_{c}(x)$, corresponding entries in the second and the
third columns are identical and equal to $\epsilon_{gc}(x)$. For each
value of $x$ considered, the corresponding value of $y_{c}(x)$ is
displayed with the accuracy necessary to reproduce the common entry in
the second and the third columns with the given precision.}
\label{TAB2}
\begin{center}
\hspace*{-0.5em}
{\small
\begin{tabular}{r@{}lcrclcl}
\hline\hline \\*[-2ex]
\multicolumn{2}{c}{$y$} && 
\multicolumn{1}{c}{$\epsilon_{g}^{(-)}(x,y)$} && 
\multicolumn{1}{c}{$\epsilon_{g}^{(+)}(x,y)$} && 
\multicolumn{1}{c}{$\epsilon_{u}(x,y)$} \\*[1ex]
\hline \\*[-0.75ex]
\multicolumn{8}{c}{$x=0.01$} \\*[2ex]
$y\to-\infty$ &&& 
\multicolumn{1}{l}{$-1+2/|y|$} &&
\multicolumn{1}{c}{---} && 
\multicolumn{1}{c}{---} \\
$-100$ &&& 
$-1+1.6054\times10^{-5}$ &&
\multicolumn{1}{c}{---} && 
\multicolumn{1}{c}{---} \\
$-10$ &&& 
$-1+1.6080\times10^{-5}$ &&
\multicolumn{1}{c}{---} && 
\multicolumn{1}{c}{---} \\
$-1$ &&& 
$-1+1.6082\times10^{-5}$ &&
$\phantom{-}1$ (exact) && 
\multicolumn{1}{c}{---} \\
1 &&& 
$-1+1.6083\times10^{-5}$ &&
$\phantom{-}1-8.1723\times10^{-5}$ && 
$\phantom{-}1$ (exact) \\
10 &&& 
$-1+1.6086\times10^{-5}$ &&
$\phantom{-}1-4.9876\times10^{-3}$ && 
$\phantom{-}1-4.9875\times10^{-3}$ \\
100 &&& 
$-1+1.6112\times10^{-5}$ &&
\multicolumn{1}{r}{$6.0000\times10^{-1}$} && 
\multicolumn{1}{r}{$6.0000\times10^{-1}$} \\
1\,000 &&& 
$-1+1.6381\times10^{-5}$ &&
$-1+7.6923\times10^{-2}$ &&
$-1+7.6923\times10^{-2}$ \\
10\,000 &&& 
$-1+1.9939\times10^{-5}$ &&
$-1+7.9219\times10^{-4}$ &&
$-1+8.0687\times10^{-4}$ \\
$y_{c}(0.01)=25\,401$ & $.358\,108\,598...$ && 
$-1+5.6189\times10^{-5}$ && 
$-1+5.6189\times10^{-5}$ && 
$-1+1.5048\times10^{-4}$ \\
100\,000 &&& 
\multicolumn{1}{c}{---} &&
\multicolumn{1}{c}{---} &&
$-1+2.3836\times10^{-5}$ \\
$y\to\infty$ &&& 
\multicolumn{1}{c}{---} &&
\multicolumn{1}{c}{---} &&
\multicolumn{1}{l}{$-1+2/y$} \\*[2ex]
\multicolumn{8}{c}{$x=0.5$} \\*[2ex]
$y\to-\infty$ &&& 
\multicolumn{1}{l}{$-1+2/|y|$} &&
\multicolumn{1}{c}{---} &&
\multicolumn{1}{c}{---} \\
$-100$ &&& 
$-1+9.6266\times10^{-3}$ &&
\multicolumn{1}{c}{---} && 
\multicolumn{1}{c}{---} \\
$-10$ &&& 
$-1+2.8822\times10^{-2}$ &&
\multicolumn{1}{c}{---} && 
\multicolumn{1}{c}{---} \\
$-1$ &&& 
$-1+3.9225\times10^{-2}$ &&
\multicolumn{1}{l}{$\phantom{-}1$ (exact)} && 
\multicolumn{1}{c}{---} \\
1 &&& 
$-1+4.3115\times10^{-2}$ &&
\multicolumn{1}{r}{$7.9970\times10^{-1}$} &&
\multicolumn{1}{l}{$\phantom{-}1$ (exact)} \\
$y_{c}(0.5)=9$ & $.436\,540\,350\,268...$ && 
\multicolumn{1}{r}{$-8.6525\times10^{-1}$} && 
\multicolumn{1}{r}{$-8.6525\times10^{-1}$} && 
\multicolumn{1}{r}{$-6.2449\times10^{-1}$} \\
10 &&& 
\multicolumn{1}{c}{---} && 
\multicolumn{1}{c}{---} &&
\multicolumn{1}{r}{$-6.5311\times10^{-1}$} \\
100 &&&
\multicolumn{1}{c}{---} && 
\multicolumn{1}{c}{---} &&
$-1+2.1489\times10^{-2}$ \\
$y\to\infty$ &&&
\multicolumn{1}{c}{---} && 
\multicolumn{1}{c}{---} &&
\multicolumn{1}{l}{$-1+2/y$} \\*[2ex]
\multicolumn{8}{c}{$x=1.5$} \\*[2ex]
$y\to-\infty$ &&&
\multicolumn{1}{l}{$-1+2/|y|$} &&
\multicolumn{1}{c}{---} && 
\multicolumn{1}{c}{---} \\
$-100$ &&& 
$-1+1.5460\times10^{-2}$ &&
\multicolumn{1}{c}{---} && 
\multicolumn{1}{c}{---} \\
$-10$ &&& 
$-1+9.4260\times10^{-2}$ &&
\multicolumn{1}{c}{---} && 
\multicolumn{1}{c}{---} \\
$-1$ &&& 
\multicolumn{1}{r}{$-7.0313\times10^{-1}$} &&
\multicolumn{1}{l}{$\phantom{-}1$ (exact)} && 
\multicolumn{1}{c}{---} \\
$y_{c}(1.5)=0$ & $.333\,896\,179\,26...$ &&
\multicolumn{1}{r}{$-1.6277\times10^{-1}$} && 
\multicolumn{1}{r}{$-1.6277\times10^{-1}$} &&
\multicolumn{1}{c}{---} \\
$y_{c}(1.5)<y<1$ &&&
\multicolumn{1}{c}{---} &&
\multicolumn{1}{c}{---} &&
\multicolumn{1}{c}{---} \\
1 &&& 
\multicolumn{1}{c}{---} &&
\multicolumn{1}{c}{---} && 
\multicolumn{1}{l}{$\phantom{-}1$ (exact)} \\
10 &&& 
\multicolumn{1}{c}{---} && 
\multicolumn{1}{c}{---} &&
\multicolumn{1}{r}{$-7.8507\times10^{-1}$} \\
100 &&&
\multicolumn{1}{c}{---} && 
\multicolumn{1}{c}{---} &&
$-1+2.0170\times10^{-2}$ \\
$y\to\infty$ &&&
\multicolumn{1}{c}{---} && 
\multicolumn{1}{c}{---} &&
\multicolumn{1}{l}{$-1+2/y$} \\*[2ex]
\hline\hline
\end{tabular}
}
\end{center}
\end{table}

Using Table \ref{TAB1}, one may establish which bound-state energy
eigenvalues exist for a given set of physical parameters
characterizing the particle ($m$) and the potentials ($R$,
$\varkappa$, $\kappa$). There are two extremes. The first one occurs
if $m$ and $R$ are such that
\begin{subequations}
\begin{equation}
\frac{\hbar}{mcR}>x_{c}
\label{6.54a}
\end{equation}
[here and then in Eqs.\ (\ref{6.55a}) and (\ref{6.57a}),
$x_{c}=1.198\,076...$ is the root to the equation $y_{c}(x_{c})=1$]
and if simultaneously $R$, $\varkappa$ and $\kappa$ are such that
\begin{equation}
y_{c}\left(\frac{\hbar}{mcR}\right)<(\varkappa-\kappa)R
\leqslant(\varkappa+\kappa)R<1.
\label{6.54b}
\end{equation}
\label{6.54}%
\end{subequations}
Then the discrete part of the particle's energy spectrum is seen to be
empty. The other extreme occurs if $m$ and $R$ are such that
\begin{subequations}
\begin{equation}
\frac{\hbar}{mcR}<x_{c}
\label{6.55a}
\end{equation}
and if simultaneously $R$, $\varkappa$ and $\kappa$ are such that
\begin{equation}
1\leqslant(\varkappa-\kappa)R\leqslant(\varkappa+\kappa)R
<y_{c}\left(\frac{\hbar}{mcR}\right).
\label{6.55b}
\end{equation}
\label{6.55}%
\end{subequations}
Then the bound-state part of the particle's energy spectrum consists
of six eigenenergies
\begin{subequations}
\begin{align}
E_{\pm g}^{(+)} & =mc^{2}\epsilon_{g}^{(+)}
\left(\frac{\hbar}{mcR},(\varkappa\pm\kappa)R\right),
\label{6.56a}
\end{align}
\begin{align}
E_{\pm g}^{(-)} & =mc^{2}\epsilon_{g}^{(-)}
\left(\frac{\hbar}{mcR},(\varkappa\pm\kappa)R\right)
\label{6.56b}
\end{align}
and
\begin{align}
E_{\pm u} & =mc^{2}
\epsilon_{u}\left(\frac{\hbar}{mcR},(\varkappa\pm\kappa)R\right)
\label{6.56c}
\end{align}
\label{6.56}%
\end{subequations}
(for $\kappa=0$ the degeneracies $E_{+g}^{(+)}=E_{-g}^{(+)}$,
$E_{+g}^{(-)}=E_{-g}^{(-)}$ and $E_{+u}=E_{-u}$ are seen to occur).
For the remaining possible combinations of $\hbar/mcR$ and
$(\varkappa-\kappa)R\leqslant(\varkappa+\kappa)R$ a variety of
intermediate cases arises. For instance, if
\begin{subequations}
\begin{equation}
\frac{\hbar}{mcR}>x_{c}
\label{6.57a}
\end{equation}
and
\begin{equation}
y_{c}\left(\frac{\hbar}{mcR}\right)<(\varkappa-\kappa)R<1
\leqslant(\varkappa+\kappa)R,
\label{6.57b}
\end{equation}
\label{6.57}%
\end{subequations}
then there is only one bound-state with eigenenergy
\begin{equation}
E_{+u}=mc^{2}
\epsilon_{u}\left(\frac{\hbar}{mcR},(\varkappa+\kappa)R\right).
\label{6.58}
\end{equation}

Next, we shall consider the question of determining the signatures
$\Delta_{a}$ of the individual eigenstates. It appears that
$\Delta_{a}$ may be correlated with the sign of the derivative
$[\partial\epsilon_{a}(\hbar/mcR,y)/\partial
y]_{y=(\varkappa\pm\kappa)R}$, i.e., with the sign of the slope of the
corresponding curve $\epsilon_{a}(x,y)$ in Fig.\ \ref{FIG3}. Indeed,
it follows from Eqs.\ (\ref{6.41}) that
\begin{equation}
\lambda_{a}(E)\equiv\lambda_{a}\big(E,(\varkappa\pm\kappa)R\big)
=\frac{\hbar}{2mcR}[(\varkappa\pm\kappa)R]
\sqrt{\frac{mc^{2}+E}{mc^{2}-E}}-1+\sigma_{a}\frac{c\hbar}{R}
\frac{\mathrm{e}^{-\sqrt{(mc^{2})^{2}-E^{2}}(R/c\hbar)}}
{\sqrt{(mc^{2})^{2}-E^{2}}},
\label{6.59}
\end{equation}
with $\sigma_{a}=+1$ for the $g$ states and $\sigma_{a}=-1$ for the
$u$ states. Now, it is an exercise in elementary calculus to prove
that if $x=x_{0}(b)$ is a root to the algebraic equation $F(x,b)=0$,
in which $x$ is a variable and $b$ is a parameter, then it holds that
\begin{equation}
\left[\frac{\partial F(x,b)}{\partial x}\right]_{x=x_{0}(b)} 
=-\left[\frac{\partial F(x,b)}{\partial b}\right]_{x=x_{0}(b)}
\left[\frac{\mathrm{d}x_{0}(b)}{\mathrm{d}b}\right]^{-1}
\label{6.60}
\end{equation}
(notice the minus sign in front of the right-hand side). On employing
the lemma (\ref{6.60}) and Eqs.\ (\ref{6.45}), from Eq.\ (\ref{6.59})
we deduce that
\begin{align}
\left[\frac{\partial\lambda_{a}\big(E,(\varkappa\pm\kappa)R\big)}
{\partial E}\right]_{E=E_{a}}
=-\frac{\hbar}{2m^{2}c^{3}R}
\sqrt{\frac{mc^{2}+E_{a}}{mc^{2}-E_{a}}}
\left[\frac{\partial\epsilon_{a}(\hbar/mcR,y)}
{\partial y}\right]_{y=(\varkappa\pm\kappa)R}^{-1},
\label{6.61}
\end{align}
from which, by virtue of Eq.\ (\ref{4.26}), it follows that
\begin{equation}
\Delta_{a}=-\sgn\left[\frac{\partial\epsilon_{a}(\hbar/mcR,y)}
{\partial y}\right]_{y=(\varkappa\pm\kappa)R}^{-1}.
\label{6.62}
\end{equation}
In conclusion, one has
\begin{equation}
\Delta_{a}=
\left\{
\begin{array}{rcl}
+1 && \textrm{for states with $\epsilon_{a}=\epsilon_{g}^{(+)}$ or
$\epsilon_{a}=\epsilon_{u}$} \\*[1ex]
0 && \textrm{for states with $\epsilon_{a}=\epsilon_{gc}$} \\*[1ex]
-1 && \textrm{for states with $\epsilon_{a}=\epsilon_{g}^{(-)}$}.
\end{array}
\right.
\label{6.63}
\end{equation}

It remains to comment on the eigenfunctions
$\Psi_{a}(\boldsymbol{r})$. If $\Delta_{a}=0$ (i.e., if
$\epsilon_{a}=\epsilon_{gc}$), the corresponding eigenfunction
$\Psi_{gc}(\boldsymbol{r})$ is an arbitrary nonzero multiple of the
Sturmian function displayed in Eq.\ (\ref{6.43a}), with $k=k_{gc}$ and
$\varepsilon=\varepsilon_{gc}$. If $\Delta_{a}=\pm1$, the normalized
[in the sense of Eq.\ (\ref{3.21})] eigenfunctions arise after one
combines Eqs.\ (\ref{4.21}) and (\ref{4.25}) with Eqs.\ (\ref{6.43})
and with the relation
\begin{equation}
\left[\frac{\partial\lambda_{a}(E)}{\partial E}\right]_{E=E_{a}}
=\frac{1}{2c\hbar\varepsilon_{a}k_{a}}
\left[\left(1+\sigma_{a}\mathrm{e}^{-k_{a}R}\right)
+\varepsilon_{a}^{2}\left(1-\sigma_{a}\mathrm{e}^{-k_{a}R}
-\sigma_{a}\frac{2\mathrm{e}^{-k_{a}R}}{k_{a}R}\right)\right],
\label{6.64}
\end{equation}
which follows once Eq.\ (\ref{6.59}) is differentiated with respect to
$E$ and then the result is simplified with the aid of Eqs.\
(\ref{6.44}) and (\ref{2.4}).

Our considerations would be incomplete without saying a few words
about the nonrelativistic limits of the energy eigenvalues.
Mathematically, the nonrelativistic regime is approached by imposing
the constraint
\begin{equation}
\frac{\hbar}{mcR}\ll1.
\label{6.65}
\end{equation}
Hence, upon retaining two leading terms in each of the truncated
series displayed in Eqs.\ (\ref{6.48}) and (\ref{6.51}), one finds the
following approximate expressions for these energy levels which are
located in the vicinity of the rest-energy threshold $mc^{2}$:
\begin{subequations}
\begin{align}
E_{\pm g}^{(+)} & \simeq mc^{2}-\frac{\hbar^{2}}{2mR^{2}}
\left[(\varkappa\pm\kappa)R
+W_{0}\big(\mathrm{e}^{-(\varkappa\pm\kappa)R}\big)\right]^{2}
\qquad [(\varkappa\pm\kappa)R\geqslant-1],
\label{6.66a}
\\
E_{\pm u} & \simeq mc^{2}-\frac{\hbar^{2}}{2mR^{2}}
\left[(\varkappa\pm\kappa)R
+W_{0}\big(\!-\!\mathrm{e}^{-(\varkappa\pm\kappa)R}\big)\right]^{2}
\qquad [(\varkappa\pm\kappa)R\geqslant1],
\label{6.66b}
\end{align}
\label{6.66}%
\end{subequations}
where, we recall, $W_{0}(z)$ is the principal branch of the Lambert
function. We have verified that the above formulas agree with those we
would get if our considerations were nonrelativistic from the
beginning.
%
%\newpage
%
\section{Conclusions}
\label{VII}
\setcounter{equation}{0}
In the previous pages, we have presented the basics of the
mathematical model for a Dirac particle bound by a set of spatially
distributed zero-range potentials. Although the applications presented
in Sec.\ \ref{VI} have been limited to only the simplest one- and
two-center systems, the developed formalism may find applications in
modeling Dirac fermions interacting with multicenter systems such as
large biomolecules, chains, lattices, and crystals, either perfect or
with structural defects.

There are several directions in which the model might be further
developed. To study systems subjected to external static electric or
magnetic fields, a suitable variant of the Rayleigh--Schr{\"o}dinger
perturbation theory should be constructed. Another challenge would be
to extend the formalism to systems involving potential centers with
internal degrees of freedom. It would be also desirable to make it
applicable to description of time-dependent processes.
\section*{Acknowledgments}
I thank Professor Sergey Leble for valuable discussions. Access to
Mathematica 12.3 through a computational grant at CI TASK is
acknowledged.
%
%\newpage
%
\appendix
\section*{Appendix: A proof of the zero-flux relation (\ref{2.12})}
\label{A}
\setcounter{equation}{0}
\setcounter{section}{1}
On employing Eq.\ (\ref{2.13}), the integral that stands on the
left-hand side of Eq.\ (\ref{2.12}) takes the form
\begin{equation}
\oint_{\mathcal{S}_{n}}\mathrm{d}^{2}\boldsymbol{\rho}_{n}\:
\boldsymbol{\mu}_{n}
\cdot c\Psi_{a}^{\dag}(\boldsymbol{r}_{n}+\boldsymbol{\rho}_{n})
\boldsymbol{\alpha}\Psi_{a}(\boldsymbol{r}_{n}+\boldsymbol{\rho}_{n}).
\label{A.1}
\end{equation}
The meaning of all symbols appearing in Eq.\ (\ref{A.1}) is the same
as in Eqs.\ (\ref{2.12}) and (\ref{2.13}), except that for convenience
we have abbreviated
$\boldsymbol{\mu}_{n}(\boldsymbol{r}_{n}+\boldsymbol{\rho}_{n})
=\boldsymbol{\rho}_{n}/\rho$ to $\boldsymbol{\mu}_{n}$. Since the
infinitesimal element $\mathrm{d}^{2}\boldsymbol{\rho}_{n}$ of the
spherical surface $\mathcal{S}_{n}$ is
\begin{equation}
\mathrm{d}^{2}\boldsymbol{\rho}_{n}
=\rho^{2}\mathrm{d}^{2}\boldsymbol{\mu}_{n},
\label{A.2}
\end{equation}
where $\mathrm{d}^{2}\boldsymbol{\mu}_{n}$ is the infinitesimal solid
angle around the direction of the unit vector $\boldsymbol{\mu}_{n}$
and with its apex at $\boldsymbol{r}_{n}$, and since
$\rho\boldsymbol{\mu}_{n}=\boldsymbol{\rho}_{n}$, the integral
(\ref{A.1}) may be rewritten as
\begin{equation}
\oint_{4\pi}\mathrm{d}^{2}\boldsymbol{\mu}_{n}\:\underbrace{c\rho
\Psi_{a}^{\dag}(\boldsymbol{r}_{n}+\boldsymbol{\rho}_{n})
\boldsymbol{\rho}_{n}\cdot\boldsymbol{\alpha}
\Psi_{a}(\boldsymbol{r}_{n}+\boldsymbol{\rho}_{n})}_{F_{a}
(\boldsymbol{\rho}_{n})}.
\label{A.3}
\end{equation}
By virtue of Eqs.\ (\ref{3.6}) and (\ref{3.7}), the integrand in Eq.\
(\ref{A.3}) may be cast into the form
\begin{equation}
F_{a}(\boldsymbol{\rho}_{n})=2c\Imag\big[\rho
\Psi_{a}^{\dag}(\boldsymbol{r}_{n}+\boldsymbol{\rho}_{n})
\mathrm{i}\boldsymbol{\rho}_{n}\cdot\boldsymbol{\alpha}^{(+)}
\Psi_{a}(\boldsymbol{r}_{n}+\boldsymbol{\rho}_{n})\big].
\label{A.4}
\end{equation}
Now, the matrix $\frac{\hbar}{2mc}\rho\mathcal{K}_{n}^{(+)}
+\varepsilon_{a}k_{a}^{-1}\beta^{(+)}$ is Hermitian (we remind that
$\varepsilon_{a}$ and $k_{a}$ are real), and consequently it holds
that
\begin{equation}
2c\Imag\left\{\rho
\Psi_{a}^{\dag}(\boldsymbol{r}_{n}+\boldsymbol{\rho}_{n})
\left[\frac{\hbar}{2mc}\rho\mathcal{K}_{n}^{(+)}
+\varepsilon_{a}k_{a}^{-1}\beta^{(+)}\right]
\Psi_{a}(\boldsymbol{r}_{n}+\boldsymbol{\rho}_{n})\right\}=0.
\label{A.5}
\end{equation}
This implies that Eq.\ (\ref{A.4}) may be equivalently written as
\begin{equation}
F_{a}(\boldsymbol{\rho}_{n})=2c\Imag\left\{\rho
\Psi_{a}^{\dag}(\boldsymbol{r}_{n}+\boldsymbol{\rho}_{n})
\left[\mathrm{i}\boldsymbol{\rho}_{n}\cdot\boldsymbol{\alpha}^{(+)}
+\frac{\hbar}{2mc}\rho\mathcal{K}_{n}^{(+)}
+\varepsilon_{a}k_{a}^{-1}\beta^{(+)}\right]
\Psi_{a}(\boldsymbol{r}_{n}+\boldsymbol{\rho}_{n})\right\}.
\label{A.6}
\end{equation}
Since the matrix that stands between the square brackets obeys
$[\ldots]=\beta^{(+)}[\ldots]$ and since $\beta^{(+)}$ is Hermitian,
it is possible to transform Eq.\ (\ref{A.6}) into
\begin{equation}
F_{a}(\boldsymbol{\rho}_{n})=2c\Imag\left\{\big[\rho\beta^{(+)}
\Psi_{a}(\boldsymbol{r}_{n}+\boldsymbol{\rho}_{n})\big]^{\dag}
\left[\mathrm{i}\boldsymbol{\rho}_{n}\cdot\boldsymbol{\alpha}^{(+)}
+\frac{\hbar}{2mc}\rho\mathcal{K}_{n}^{(+)}
+\varepsilon_{a}k_{a}^{-1}\beta^{(+)}\right]
\Psi_{a}(\boldsymbol{r}_{n}+\boldsymbol{\rho}_{n})\right\}.
\label{A.7}
\end{equation}
It follows from Eqs.\ (\ref{2.2}) and (\ref{2.3a}) that the limit of
$\rho\beta^{(+)}\Psi_{a}(\boldsymbol{r}_{n}+\boldsymbol{\rho}_{n})$ as
$\rho\to0$ is finite. On the other hand, by virtue of the constraints
(\ref{2.7}), one has
\begin{equation}
\lim_{\rho\to0}
\left[\mathrm{i}\boldsymbol{\rho}_{n}\cdot\boldsymbol{\alpha}^{(+)}
+\frac{\hbar}{2mc}\rho\mathcal{K}_{n}^{(+)}
+\varepsilon_{a}k_{a}^{-1}\beta^{(+)}\right]
\Psi_{a}(\boldsymbol{r}_{n}+\boldsymbol{\rho}_{n})=0.
\label{A.8}
\end{equation}
This implies that in the limit $\rho\to0$ the integrand in Eq.\
(\ref{A.3}) vanishes. Remembering the equivalence of the integrals in
Eqs.\ (\ref{A.3}) and (\ref{A.1}), we thus obtain
\begin{equation}
\lim_{\rho\to0}
\oint_{\mathcal{S}_{n}}\mathrm{d}^{2}\boldsymbol{\rho}_{n}\:
\boldsymbol{\mu}_{n}(\boldsymbol{r}_{n}+\boldsymbol{\rho}_{n})
\cdot c\Psi_{a}^{\dag}(\boldsymbol{r}_{n}+\boldsymbol{\rho}_{n})
\boldsymbol{\alpha}\Psi_{a}(\boldsymbol{r}_{n}+\boldsymbol{\rho}_{n})
=0,
\label{A.9}
\end{equation}
which, after being combined with Eq.\ (\ref{2.13}), is seen to
coincide with Eq.\ (\ref{2.12}).
%
%\newpage
%

%
\end{document}